\documentstyle[preprint,tighten,aps]{revtex}
\begin{document}
\draft
\title{Fokker-Planck description of the transfer matrix limiting 
distribution in the scattering approach to quantum transport} 
\author{Dirk Endesfelder}
\address{Oxford University,
Theoretical Physics, \\
1 Keble Road, \\
United Kingdom}
\date{\today}
\maketitle
\begin{abstract}
The scattering approach to quantum transport through a disordered 
quasi-one-dimensional conductor in the insulating regime is discussed in 
terms of its transfer matrix $\bbox{T}$. A model of $N$ one-dimensional 
wires which are coupled by random hopping matrix elements is compared 
with the transfer matrix model of Mello and Tomsovic. We derive and 
discuss the complete Fokker-Planck equation which describes 
the evolution of the probability distribution of $\bbox{TT}^{\dagger}$ 
with system length in the insulating regime. It is demonstrated that the
eigenvalues of $\ln\bbox{TT}^{\dagger}$ have a multivariate Gaussian 
limiting probability distribution. The parameters of the distribution 
are expressed in terms of averages over the stationary distribution of 
the eigenvectors of $\bbox{TT}^{\dagger}$. We compare the general form
of the limiting distribution with results of random matrix theory and
the Dorokhov-Mello-Pereyra-Kumar equation. 
\end{abstract}
\pacs{PACS numbers: 72.10.Bg, 05.60.+w, 72.15.Rn, 73.50.Bk}
\section{Introduction}              
The statistical properties of phase coherent quantum transport in 
mesoscopic systems have received increasing attention during the past
few years \cite{MesoscopicRev}. A variety of low temperature 
transport quantities of conductors which are coupled to ideal leads can
be expressed in terms of their scattering properties 
\cite{Buettiker90a,Beenakker91a,Beenakker92a}. Hence, their statistics
may be studied in terms of the probability distribution of the 
scattering matrix. Several distinct  approaches, including random matrix
theory (RMT) \cite{Imry86a,Muttalib87a,Beenakker93c,Beenakker93a}, 
Fokker-Planck (FP) equations 
\cite{Mello88a,Mello88c,Hueffmann90a,Mello91b,Macedo92a},
supersymmetry methods \cite{Iida90b,Iida90a,Altland91a}, and 
diagrammatic techniques \cite{Lee85a,Altshuler85a} have been employed.
This led to considerable progress in the understanding of 
quasi-one-dimensional wires 
\cite{Chalker93c,Macedo94a,Beenakker93b,Beenakker94a,Caselle95a} whose 
width is of the order of the mean free path which implies a 
structureless cross-section since no transverse diffusion takes place.
The mean and the variance of the conductance of quasi-one-dimensional
wires are now known for all length scales from the metallic to the 
localized regime \cite{Zirnbauer92a,Mirlin94a,Frahm95a}. The 
generalization of these results beyond the quasi-one-dimensional regime
is of considerable interest and has been the subject of some recent work 
\cite{Mello91c,Mello92a,Chalker93a,Nazarov94a}. 
Having this goal in mind, we focus on wires in the localized regime 
which are still quasi-one-dimensional in the sense that they are much 
longer than wide but which are not structureless in the transverse 
direction. In this regime the FP description simplifies considerably and
progress is possible.\\
This paper, which is the first of a series of two, deals mainly with the 
technical aspects of the problem and compares the general result which 
is obtained with previous results from random-matrix theory  and 
the Dorokhov-Mello-Pereyra-Kumar (DMPK) \cite{Dorokhov82a,Mello88c} 
equation. It has some overlap with the pioneering work of Dorokhov 
\cite{Dorokhov83a} but goes beyond it by generalizing the derivation of 
the transfer matrix limiting distribution for the one-dimensional wire
by Kree and Schmid \cite{Kree81a} to the quasi-one-dimensional case. A
preliminary account of the results which are presented here has been 
given in \cite{Endesfelder93a}. In the second paper \cite{Endesfelder96c} 
we will investigate a model in which forward scattering is much stronger 
than backward scattering. We use the ratio of backward to forward 
scattering strength as a small expansion parameter and calculate the 
limiting distribution in the lowest two orders.

The transfer matrix transforms the amplitudes of the propagating wave 
modes (open channels) at the Fermi energy in the left lead into the
amplitudes of the right lead. A convenient parametrization for 
conductors with time-reversal invariance and with no spin-orbit 
scattering is the polar decomposition \cite{Dorokhov83a,Mello91a} 
\begin{equation}
\label{ParT1}
\bbox{T}=\left(\begin{array}{cc} \bbox{u} & \bbox{0} \\
                   \bbox{0} & \bbox{u}^{*} \end{array}\right)
\left(\begin{array}{cc} 
      \sqrt{\bbox{1}+\bbox{\lambda}} & \sqrt{\bbox{\lambda}} \\
      \sqrt{\bbox{\lambda}} & \sqrt{\bbox{1}+\bbox{\lambda}} 
      \end{array}\right)
\left(\begin{array}{cc} \bbox{v} & \bbox{0} \\
                        \bbox{0} & \bbox{v}^{*} \end{array}\right)  ,
\end{equation}
where  $\bbox{u}$, $\bbox{v}$ are unitary $N\times N$ matrices and 
$\bbox{\lambda}$ is diagonal with real and positive diagonal elements 
$\lambda_{i}$. The two-terminal conductance in units of $e^{2}/h$ is
$g=\sum_{i}{\cal T}_{i}$ where ${\cal T}_{i}=1/(1+\lambda_{i})$ are the 
transmission eigenvalues of  $\bbox{tt}^{\dagger}$ and $\bbox{t}= 
\bbox{u}(\bbox{1}/\sqrt{\bbox{1}+\bbox{\lambda}})\bbox{v}$ is the
transmission matrix.
 
The transfer matrix of two samples which are joined together is the
product of the transfer matrices of the individual samples. Building up 
a long wire by combining short samples, thus leads to a transfer matrix 
which is a product of a large number of random matrices. The 
eigenvalues of $\ln(\bbox{TT}^{\dagger})/2L$ come in pairs $(\alpha_{m}
(L),-\alpha_{m}(L))$ where $1+2\lambda_{m}\equiv\cosh(2\alpha_{m}L)
\equiv\cosh\Gamma_{m}$ and $L$ is the system length. The corresponding 
eigenvectors are $(\vec{u}^{T}_{m},\vec{u}^{*\: T}_{m})^{T}/\sqrt{2}$ 
and $(\vec{u}^{T}_{m},-\vec{u}^{*\: T}_{m})^{T}/\sqrt{2}$ where 
$\vec{u}_{m}$ is the $m^{th}$ column vector of $\bbox{u}$.

From Oseledec's theorem \cite{Oseledec68a} for 
random matrix products it is known that the $\alpha_{m}(L)$ are 
self-averaging and distinct if $L$ goes to infinity. The limiting values
$\alpha^{\infty}_{m}\equiv\lim_{L\to\infty}\alpha_{m}(L)$ are the 
Lyapunov exponents. They characterize the rate of exponential growth of 
the $\lambda_{m}$ with system length. Furthermore there are central 
limit theorems \cite{Tutubalin68a,Virtser70a} which show that $\bbox{u}$
and $\bbox{v}$ have stationary distributions and that the quantities 
$(\Gamma_{m}-2\alpha^{\infty}_{m}L)/\sqrt{L}$ have Gaussian limiting
distributions if $L$ goes to infinity. Oseledec's theorem implies that
the $\Gamma_{m}$ can be ordered as $1\ll\Gamma_{1}\ll\Gamma_{2}
\cdots\ll\Gamma_{N}$ if $2\alpha_{1}^{\infty}L\gg 1$. In this regime
$g=\sum_{m}2/(2+\cosh\Gamma_{m})\approx 4\exp(-\Gamma_{1})=4\exp(-2
\alpha_{1}L)$ and the sample is strongly insulating. The decay length 
$\xi=1/2\alpha^{\infty}_{1}$ of the typical conductance is usually 
identified with the localization length. Johnston and Kunz 
\cite{Johnston83a} applied the central limit theorems to the Anderson
model. We derive the Gaussian limiting distribution within a FP
approach, which will establish a link between the parameters of the 
limiting distribution and the stationary distribution of $\bbox{u}$. 

The determination of the Lyapunov exponents of random matrix products
is a problem which arises often in the context of disordered systems. 
At present there is no powerful method to calculate them analytically. 
Only special cases like sparse random matrices have been solved 
\cite{Cook90a}. The weak disorder expansions which have been developped 
cannot be succesfully applied to quasi-one-dimensional conductors 
because of the problem of degenerate eigenvalues 
\cite{Derrida87a,Zanon88a,Markos89a,Markos93c}. The full limiting 
distribution has been mainly studied in numerical simulations 
\cite{Markos93a,Markos94a}. Apart from the numerical data there are only 
two analytical approaches which make quantitative predictions, RMT 
\cite{Imry86a,Muttalib87a,Pichard91a} and the DMPK equation 
\cite{Beenakker94a,Caselle95a}.
  
The RMT ensemble maximizes the information entropy of the transfer 
matrix probability distribution subject to the constraint of a given 
density $\langle\rho(\Gamma)\rangle\equiv\langle\sum_{m}\delta(\Gamma-
\Gamma_{m})\rangle$. As a consequence, $\bbox{\Gamma}$ and the unitary
matrices are stochastically independent. The unitary matrices are 
isotropically distributed and the probability distribution of 
$\bbox{\Gamma}$ has the form  
\begin{equation}
\label{RMTp}
p(\bbox{\Gamma})={\cal N}^{-1}\exp\{-\beta H(\bbox{\Gamma})\},
\end{equation}
where
\begin{equation}
\label{RMTHam}
H(\bbox{\Gamma}) =  -\sum_{m<n}\ln |\cosh\Gamma_{m}-\cosh\Gamma_{n}|
-\frac{1}{\beta}\sum_{m}\ln(\sinh\Gamma_{m})+\sum_{m} V(\Gamma_{m})
\end{equation}
and $\cal{N}$ is a normalization factor. The parameter $\beta$ is 
determined by the symmetry of the transfer matrix ensemble. The 
orthogonal $(\beta=1)$, unitary $(\beta=2)$, and symplectic $(\beta=4)$
ensembles correpond to conductors with time reversal symmetry, broken 
time reversal symmetry, and strong spin-orbit scattering respectively. 
The potential $V(\lambda)$ has to be determined from $\langle\rho(
\Gamma)\rangle$. RMT is known to be a good but not exact description of 
quasi-one-dimensional conductors without transverse structure 
\cite{Beenakker93c}. For such conductors and large $N$, $\langle\rho(
\Gamma)\rangle\approx Nl/2L$ if $0\le\Gamma<2L/l$ and $\langle\rho(
\Gamma)\rangle\approx 0$ if $\Gamma>2L/l$ where $l$ is the mean free 
path. The resulting potential is quadratic $V(\Gamma)\approx Nl
\Gamma^{2}/4L$ \cite{Pichard90a}. In the insulating regime where $1\ll
\Gamma_{1}\ll\Gamma_{2}\cdots\ll\Gamma_{N}$ the Hamiltonian 
(\ref{RMTHam}) simplifies, since $\ln(\sinh\Gamma_{m})\approx\Gamma_{m}$
and $\ln |\cosh\Gamma_{m}-\cosh\Gamma_{n}|\approx\Gamma_{n}$ if $m<n$. 
This leads to the Gaussian probability distribution 
\begin{equation}
\label{RMTGauss}
p(\bbox{\Gamma}) =  
\prod_{m=1}^{N}\frac{1}{\sqrt{2\pi\sigma^{2}}}\exp\left\{-\frac{(
\Gamma_{m}-\langle\Gamma_{m}\rangle)^{2}}{2\sigma^{2}}\right\} 
\end{equation} 
where $\langle\Gamma_{m}\rangle=(m-1+1/\beta)2L/(l N)$ and 
$\sigma^{2}=2L/(\beta l N)$. Note that any potential must have the form
$V(\Gamma)\approx a\Gamma+b\Gamma^{2}$ if $L\gg\xi$ and $\Gamma\gg 1$ in 
order to be consistent with the Gaussian limiting distribution. This 
implies always equidistant mean values $\Gamma_{m}$ and equal variances 
for the fluctuations around them. Numerical simulations of conductors 
with transverse structure show that this is in general not true 
\cite{Pichard86a,Markos93a,Markos94a}. Therefore RMT cannot describe 
such conductors.  
 
The DMPK equation 
\begin{equation}
\frac{\partial p(L;\bbox{\Gamma})}{\partial L}=
\frac{2}{l\gamma}\sum_{m=1}^{N}\frac{\partial}{\partial\Gamma_{m}}
\left(\frac{\partial p}{\partial\Gamma_{m}}+\beta p\frac{\partial
\Omega(\bbox{\Gamma})}{\partial\Gamma_{m}}\right),
\end{equation}
where $\Omega(\bbox{\Gamma})=-\sum_{m<n}\ln |(\cosh\Gamma_{m}-\cosh
\Gamma_{n})/2|-1/\beta\sum_{m}\ln |\sinh\Gamma_{m}|$ and $\gamma=
\beta N+2-\beta$, constitutes an exact description of 
quasi-one-dimensional wires without transverse structure. Its solution
\begin{eqnarray}
p(\bbox{\Gamma}) & \propto & \prod_{m<n}|\cosh\Gamma_{m}-\cosh\Gamma_{n}
|^{\beta/2}|\Gamma_{m}^{2}-\Gamma_{n}^{2}|\nonumber\\
& & \times\prod_{m}[\exp(-\gamma\Gamma_{m}^{2}l/8 L)\Gamma_{m}
(\sinh\Gamma_{m})^{1/2}]
\end{eqnarray}
in the insulating regime \cite{Beenakker94a,Caselle95a} can be as well 
approximated by a Gaussian distribution of the form (\ref{RMTGauss}) if 
$1\ll\Gamma_{1}\ll\Gamma_{2}\cdots\ll\Gamma_{N}$, where $\langle
\Gamma_{m}\rangle=[1+\beta(m-1)]2L/[l(\beta N+2-\beta)]$ and $\sigma^{2}
=4L/[l(\beta N+2-\beta)]$. Note that the mean values of RMT and of the 
DMPK equation coincide for large $N$ whereas the variances differ by a 
factor of two.

The content of the paper is organized as follows. In Sec. 
\ref{SScatModels} a Hamiltonian model of $N$ one-dimensional wires
which are coupled by random hopping matrix elements is compared to the
transfer matrix model of Mello and Tomsovic \cite{Mello91c,Mello92a}.
In Sec. \ref{SLvFPgl} we derive the FP equation which describes 
the evolution of the probability distribution of $\bbox{\Gamma}$ and 
$\bbox{u}$ with system length in the localized regime.  
In Sec. \ref{Slimdistr} we generalize the derivation for the transfer 
matrix limiting distribution of a one-dimensional wire by Kree and 
Schmid \cite{Kree81a} to the quasi-one-dimensional wire. 
A first application of this approach is presented in Sec. \ref{SECM}, 
where we investigate the Equivalent Channel Model (ECM) of Mello and 
Tomsovic. The joint probability distribution $p(L;\bbox{\Gamma})$ of 
this model is known to be identical to the distribution of the DMPK 
equation for $\beta=1$. We recover the Gaussian distribution 
(\ref{RMTGauss}) and show that the stationary distribution of 
$\bbox{u}$ is isotropic.

There are four appendices. The derivation of the FP equation in Sec. 
\ref{SLvFPgl} is based on a simplified version of the general Langevin
equations for $\bbox{\Gamma}$ and $\bbox{u}$ which are obtained in 
Appendix \ref{ALvgl}. The coefficients of the FP operator are derived 
in Appendix \ref{AFPkoeff}. In Appendix \ref{AInvMeas} we show that a 
measure for the unitary group which has been introduced in the text is 
the invariant measure. An alternative derivation of the FP equation is
presented in Appendix \ref{AaltderFP}. The summation convention is used 
throughout the whole paper. 
\section{Hamiltonian versus transfer matrix models}
\label{SScatModels}
The FP approach to disordered conductors has been pioneered
by Dorokhov \cite{Dorokhov83a}. He started from a microscopic model of
disordered coupled chains which led to a quite complicated FP equation. 
Similar models were also studied by other techniques 
\cite{Weller82a,Apel83a,Prigodin86a,Weller88a,Kasner88a}.
Recently Mello and Tomsovic proposed a class of models which was 
formulated  on the level of the transfer matrix 
\cite{Mello91c,Mello92a}. On the one hand these models lead to simpler 
FP equations but on the other hand the underlying Hamiltonian is not 
known. In this section we propose a microscopic model which is simpler 
than the one which has been used by Dorokhov and compare it to the model
class of Mello and Tomsovic.    

Consider the scattering of electrons at a quasi-1D disordered conductor 
with a $(d-1)$-dimensional cross section which is connected to perfectly 
ordered leads. The conductor consists of $N$ 1D-wires which are only 
coupled by random hopping matrix elements. It is described by the 
Hamiltonian
\begin{equation} 
\label{Mod1a}
H_{nn'}= -\delta_{nn'}\frac{\hbar^2}{2m_{0}} \partial_x^2 +V_{nn'}(x), 
\end{equation}
where $V_{nn'}(x)$ is real and symmetric in its indices and $n=1,\cdots,
N$. The potential $V_{nn'}(x)$ is zero in the leads and stochastic in 
the disordered system of length $L$. It describes on-wire disorder for 
$n=n'$ and random hopping between the wires for $n\ne n'$. The 
independent matrix elements of $\bbox{V}(x)$ are chosen to be 
uncorrelated and Gaussian distributed with zero average          
\begin{eqnarray}
\label{Mod1b}
\langle V_{nn'}(x)\rangle & = & 0 \nonumber \\
\langle V_{nn'}(x) V_{mm'}(x')\rangle & = &  U_{nn'}\delta (x-x')
(\delta_{nm}\delta_{n'm'}+\delta_{nm'}\delta_{n'm})
\end{eqnarray}
where $U_{nn'}=U_{n'n}$. The special case $U_{nn'}=U/N$ can be
interpreted as a continuous                                 
one-dimensional $N$-orbital model \cite{Wegner79a}, which is connected 
to ideal leads with no exponential decaying modes.

The solution $\Psi(xn;E)$ of the scattering problem with the incoming 
waves 
\begin{equation}
\label{ScatSol}
\Psi^{in}(xn;E) = \sqrt{\frac{m_{0}}{\hbar k}} \left( a^{l}_{n} 
\exp\{ikx\}+b^{r}_{n} \exp\{-ikx\}\right)
\end{equation} 
is an eigenfunction of the Schr\"odinger equation with energy $E=
\hbar^{2}k^{2}/2m_{0}$. Its form in the left and the right lead 
respectively is
\begin{equation}
\Psi^{l/r}(xn;E) = \sqrt{\frac{m_{0}}{\hbar k}} \left( a^{l/r}_{n} 
\exp\{ikx\}+b^{l/r}_{n} \exp\{-ikx\}\right)
\end{equation} 
where the amplitudes $a^{l/r}_{n}$ and $b^{l/r}_{n}$ have been 
normalized in such a way that the probability current in x-direction 
is $j_{x}=\sum_{n} |a_{n}|^{2}-|b_{n}|^{2}$. The $\bbox{S}$-matrix 
transforms the amplitudes of the incident waves into the amplitudes of 
the scattered waves
\begin{equation} 
\label{DefS}
\begin{array}{ccccccc} 
\left(\begin{array}{c} \bbox{b}^{l}\\ \bbox{a}^{r} \end{array} \right) 
& = & \bbox{S} 
\left(\begin{array}{c} \bbox{a}^{l} \\ \bbox{b}^{r}\end{array} \right) 
& , & \bbox{S} & = &
\left(\begin{array}{cc} \bbox{r} & \bbox{t}' \\ 
                        \bbox{t} & \bbox{r}' \end{array}\right) 
\end{array} .
\end{equation}
Current conservation and time reversal invariance imply that $\bbox{S}$ 
is unitary and symmetric respectively. The transfer matrix by contrast 
transforms the amplitudes in the left lead into the amplitudes in the 
right lead
\begin{equation}
\label{DefT}
\begin{array}{ccccccc} 
\left(\begin{array}{c} \bbox{a}^{r}\\ \bbox{b}^{r} \end{array} \right) 
& = & \bbox{T} 
\left(\begin{array}{c} \bbox{a}^{l} \\ \bbox{b}^{l}\end{array} \right)
& , &  \bbox{T} = \left(\begin{array}{cc} 
\bbox{t}-\bbox{r}'\bbox{t}'^{-1}\bbox{r} & \bbox{r}'\bbox{t}'^{-1} \\
-\bbox{t}'^{-1}\bbox{r} & \bbox{t}'^{-1} \end{array}\right). 
\end{array}
\end{equation}
Here, current conservation and time reversal invariance leads to the 
form
\begin{equation}
\label{ParT2}
\bbox{T} =  \left(\begin{array}{cc} 
                  \bbox{\alpha}    & \bbox{\beta} \\ 
                  \bbox{\beta}^{*} & \bbox{\alpha}^{*} 
                  \end{array}\right) 
\end{equation}
where $\bbox{\alpha\alpha}^{\dagger}-\bbox{\beta\beta}^{\dagger}=
\bbox{1}$ \cite{Mello88a,Mello88c}. Apart from the polar decomposition 
(\ref{ParT1}) there is another useful parametrization of the transfer 
matrix which has been introduced by Mello and Tomsovic \cite{Mello92a} 
and has the form  
\begin{equation}
\label{ParT3}
\bbox{T}=\left(\begin{array}{cc} 
               \exp\bbox{\vartheta} & \bbox{0} \\
               \bbox{0} & \exp\bbox{\vartheta}^{*} 
               \end{array}\right)
\left(\begin{array}{cc} 
\sqrt{\bbox{1}+\bbox{\eta\eta^{*}}} & \bbox{\eta} \\
\bbox{\eta}^{*} & \sqrt{\bbox{1}+\bbox{\eta}^{*}\bbox{\eta}}
\end{array}\right)
\end{equation}
where $\bbox{\vartheta}$ and $\bbox{\eta}$ are complex $N\times N$ 
matrices and $\bbox{\vartheta}^{\dagger}=-\bbox{\vartheta}$ and 
$\bbox{\eta}^{T}=\bbox{\eta}$.
The wave amplitudes $\bbox{a}_{l/r}$ and $\bbox{b}_{l/r}$ fix the 
values of $\Psi(x,n;E)$ and $\partial_{x}\Psi(x,n;E)$ at the edges of 
the sample. This implies that the transfer matrix of two samples which
are matched continuously together is  
\begin{equation}
\label{Tmultiplicativity}
\bbox{T}(L+L',0)=\bbox{T}(L',L)\bbox{T}(L,0) .
\end{equation}
Hence, the transfer matrix of a sample of length $L$ can be obtained by
dividing it into short segments of length $\delta L$ and multiplying
the transfer matrices of the segments. The evolution of the transfer 
matrix with the system length is a multiplicative stochastic process.
It can be described by a Langevin equation since the model is continuous 
in the scattering direction. The Langevin equation has the form
\begin{equation}
\frac{d\bbox{T}(x,0)}{dx}=
\left(\begin{array}{ll} 
      \bbox{\gamma}^{11}(x) & \bbox{\gamma}^{12}(x) \\
      \bbox{\gamma}^{21}(x) & \bbox{\gamma}^{22}(x) \end{array}\right)
\bbox{T}(x,0)
\end{equation}
with the noise $\gamma^{ij}(x)$. The symmetries 
\begin{eqnarray}
\label{symm1}
\bbox{\gamma}^{22} & = & \bbox{\gamma}^{11\: *} \nonumber \\
\bbox{\gamma}^{21} & = & \bbox{\gamma}^{12\: *},
\end{eqnarray}
and
\begin{eqnarray}
\label{symm2}
\bbox{\gamma}^{11\:\dagger} & = & -\bbox{\gamma}^{11} \nonumber \\
\bbox{\gamma}^{12\: T} & = & \bbox{\gamma}^{12}
\end{eqnarray}
which will be derived below ensure time reversal invariance and
flux conservation. Iterative integration of the Langevin equation 
yields 
\begin{eqnarray}
\label{Texpansion}
\bbox{T}(x_{0}+\delta L,x_{0}) & = &  
\bbox{1}+\left(\begin{array}{ll} 
\bbox{\varepsilon}^{11} & \bbox{\varepsilon}^{12} \\
\bbox{\varepsilon}^{21} & \bbox{\varepsilon}^{22} 
\end{array}\right)  
\end{eqnarray}
where 
\begin{equation}
\label{T2expansion}
\bbox{\varepsilon}^{ij} = \int_{x_{0}}^{x_{0}+\delta L} dx 
\bbox{\gamma}^{ij}(x)+\int_{x_{0}}^{x_{0}+\delta L} dx \int_{x_{0}}^{x} 
dx'\bbox{\gamma}^{ik}(x)\bbox{\gamma}^{kj}(x')+\cdots . 
\end{equation}
For uncorrelated noise the first term of this expansion is of order
$(\delta L)^{1/2}$ and the second term is of order $\delta L$. 
For the derivation of the symmetries (\ref{symm1}) and (\ref{symm2}), 
however, it is convenient to start with a finite correlation length of
the noise. 
Taking $\delta x\le\delta L$ and $\delta L$ to be smaller than this 
correlation length one may 
expand $\bbox{\gamma^{ij}}(x_{0}+\delta x)=\bbox{\gamma^{ij}}(x_{0})+
\partial_{x}\bbox{\gamma}^{ij}(x_{0})\delta x+O((\delta x)^{2})$ which
leads to 
\begin{equation}
\label{T3expansion}
\bbox{\varepsilon}^{ij}=\bbox{\gamma}^{ij}(x_{0})\delta L+
                         O((\delta L)^{2}).
\end{equation}
Equation (\ref{ParT2}) enforces the symmetries $\bbox{\gamma}^{22}=
\bbox{\gamma}^{11\: *}$ and $\bbox{\gamma}^{21}=\bbox{\gamma}^{12\:
*}$. Comparing the expansion (\ref{T3expansion}) with the
parametrization (\ref{ParT3}) of the transfer matrix one finds 
\begin{eqnarray} 
\bbox{\vartheta}(x_{0}+\delta L,x_{0}) & = & \bbox{\gamma}^{11}(x_{0})
\delta L+O(\delta L^{2}) \nonumber\\
\bbox{\eta}(x_{0}+\delta L,x_{0}) & = & \bbox{\gamma}^{12}(x_{0})\delta 
L+O(\delta L^{2})
\end{eqnarray}
which implies $\bbox{\gamma}^{11\:\dagger}=-\bbox{\gamma}^{11}$ and 
$\bbox{\gamma}^{12\: T}=\bbox{\gamma}^{12}$. These symmetries remain
valid in the limit of zero correlation length. 

In the sequel we derive $\bbox{\gamma}^{11}(x_{0})$ and $\bbox{\gamma}^{
12}(x_{0})$ for the Hamiltonian model (\ref{Mod1a}). The stationary 
solution of the scattering problem obeys the Lippmann-Schwinger equation
\begin{equation}
\label{LiScGl}
\Psi(xn;E) = \Psi^{in}(xn;E)+\int\mbox{d}x_{1}\sum_{n_{1},n_{1}'} 
G^{+}_{0}(xn,x_{1}n_{1};E) V_{n_{1}n_{1}'}(x_{1}) \Psi(x_{1},n_{1}';E) ,
\end{equation}
where 
\begin{equation}
G^{+}_{0}(xn,x'n';E)=
\frac{-im_{0}\delta_{nn'}}{\hbar^{2}k(E)}\exp\{ik(E)|x-x'|\}
\end{equation}
is the free retarded Green's function. Iteration of the 
Lippmann-Schwinger yields the Born series
\begin{eqnarray}
\Psi(xn;E) & = & \Psi^{in}(xn;E) \nonumber\\
&  & +\int_{x_{0}}^{x_{0}+\delta L} \mbox{d}x_{1}\sum_{n_{1},n_{1}'} 
G^{+}_{0}(xn,x_{1}n_{1};E) V_{n_{1}n_{1}'}(x_{1})
\Psi^{in}(x_{1},n_{1}';E)+\cdots 
\end{eqnarray}
which can be translated into series for the transmission and reflection 
matrices by Eq. (\ref{ScatSol}) and Eq. (\ref{DefS}). The $n^{th}$ 
orders of these series are at least of the order $(\delta L)^{n}$ since 
they contain $n$ integrations from $x_{0}$ to $x_{0}+\delta L$. Thus, 
only the first orders can contribute to the terms of order $\delta L$ of
the expansions $\bbox{t}=\bbox{1}+\bbox{t}^{1}\delta L+\cdots\;$, 
$\bbox{r}=\bbox{r}^{1}\delta L+\cdots\;$, $\bbox{t}'=\bbox{1}+
\bbox{t}'^{\: 1}\delta L+\cdots\;$, $\bbox{r}'=\bbox{r}'^{\: 1}\delta L+
\cdots\;$. Inserting these contributions into the relations $\bbox{
\gamma}^{11}=\bbox{t}^{1}$ and $\bbox{\gamma}^{12}=\bbox{r}'^{\:1}$ 
which follow from Eq. (\ref{DefT}) and Eq. (\ref{T3expansion}) yields  
\begin{eqnarray}
\label{Tfirstorder}
\gamma^{11}_{nn'}(x_{0}) & = & \frac{-im_{0}}{\hbar^{2}k} V_{nn'}(x_{0}) 
\nonumber \\
\gamma^{12}_{nn'}(x_{0}) & = & \frac{-im_{0}\exp(-i2kx_{0})}{\hbar^{2}k} 
V_{nn'}(x_{0}) .
\end{eqnarray}
The phase $\exp(-i2kx_{0})$ which appears in $\bbox{\gamma}^{12}(x_{0})$ 
is a consequence of the transformation rule   
\begin{equation}
\bbox{T}(L+x_{0},x_{0})=
\left(\begin{array}{cc} \mbox{e}^{-ikx_{0}}\bbox{1} & \bbox{0} \\
                        \bbox{0} & \mbox{e}^{ikx_{0}}\bbox{1} 
      \end{array}\right)
\bbox{T}(L,0)
\left(\begin{array}{cc} \mbox{e}^{ikx_{0}}\bbox{1} & \bbox{0} \\
                        \bbox{0} & \mbox{e}^{-ikx_{0}}\bbox{1}
      \end{array}\right),
\end{equation}
which accounts for a shift of the disordered region by $x_{0}$.

Now we are in the position to compare the Hamiltonian model
(\ref{Mod1b}) with the transfer matrix model of Mello and Tomsovic 
\cite{Mello91c,Mello92a}. They divided a sample of length $L$ into $n=L/
\delta L$ uncorrelated scattering units with identical statistical 
properties. Specifying the first two moments of $\vartheta_{mn}$ and 
$\eta_{mn}$ for one scatterer
\begin{eqnarray}
\label{Mod2a}
\langle\vartheta_{mn}\rangle_{\delta L}=\langle\eta_{mn}
\rangle_{\delta L} & = & 0 \nonumber\\
\langle\vartheta_{mn}\vartheta_{m'n'}\rangle_{\delta L} & = & 
\kappa^{11,11}_{mn,m'n'}\nonumber\\
\langle\vartheta_{mn}\vartheta^{*}_{m'n'}\rangle_{\delta L} & = & 
\kappa^{11,22}_{mn,m'n'}\nonumber\\
\langle\eta_{mn}\eta_{m'n'}\rangle_{\delta L} & = & 
\kappa^{12,12}_{mn,m'n'}\nonumber\\
\langle\eta_{mn}\eta^{*}_{m'n'}\rangle_{\delta L} & = & 
\kappa^{12,21}_{mn,m'n'}\nonumber\\
\langle\vartheta_{mn}\eta_{m'n'}\rangle_{\delta L} & = & 
\kappa^{11,12}_{mn,m'n'} \nonumber\\
\langle\vartheta_{mn}\eta^{*}_{m'n'}\rangle_{\delta L} & = & 
\kappa^{11,21}_{mn,m'n'} 
\end{eqnarray}
and taking the continuum limit of a high number of weak scattering
units such that
\begin{equation}
\label{Mod2b}
\lim_{\delta L\to 0}\frac{1}{\delta L}\kappa^{ij,i'j'}_{mn,m'n'}= 
\sigma^{ij,i'j'}_{mn,m'n'} 
\end{equation}
and that $1/\delta L$ times higher moments gives zero in the same
limit determined completely the stochastic evolution of the transfer 
matrix. As a consequence one finds  
\begin{eqnarray}
\label{MomMod2}
[\varepsilon^{11}_{mn}] & = & 
(\sigma^{11,11}_{mk,kn}+\sigma^{12,21}_{mk,kn})/2\nonumber\\[1ex]
[\varepsilon^{12}_{mn}] & = & \sigma^{11,12}_{mk,kn}\nonumber\\[1ex] 
[\varepsilon^{ij}_{mn}\varepsilon^{i'j'}_{m'n'}] & = & 
\sigma^{ij,i'j'}_{mn,m'n'}
\end{eqnarray}
where $[\cdots]\equiv\lim_{\delta L\to 0}\langle\cdots\rangle_{
\delta L}/\delta L$. The same limit for higher moments of $\varepsilon^{
ij}_{mn}$ is zero. Mello and Tomsovic have chosen the following
simple model for one scattering unit. Assume that the 
independent matrix elements of $\bbox{\vartheta}$ and $\bbox{\eta}$
are uncorrelated and that their phases are randomly distributed. 
Averaging over the arbitrary distribution of their modulus then leads
to
\begin{eqnarray}
\label{MomMod2special}
[\varepsilon^{ij}_{mn}] & = & \delta_{ij}\delta_{mn}
\left(\frac{1}{l^{b}}-\frac{1}{l^{f}}\right)\nonumber\\[1ex]
[\varepsilon^{11}_{mn}\varepsilon^{11}_{m'n'}] 
& = & -\delta_{mn'}\delta_{nm'}\frac{1}{l^{f}_{mn}} 
\nonumber \\[1ex]
[\varepsilon^{11}_{mn}\varepsilon^{22}_{m'n'}] 
& = & \delta_{mm'}\delta_{nn'}\frac{1}{l^{f}_{mn}} \nonumber \\[1ex]
[\varepsilon^{12}_{mn}\varepsilon^{21}_{m'n'}] 
& = & \frac{\delta_{mm'}\delta_{nn'}+\delta_{mn'}\delta_{nm'}}{1+
\delta_{mn}}\frac{1}{l^{b}_{mn}} \nonumber \\[1ex]
[\varepsilon^{11}_{mn}\varepsilon^{12}_{m'n'}] 
& = & 0 \nonumber \\[1ex]
[\varepsilon^{11}_{mn}\varepsilon^{21}_{m'n'}] 
& = & 0 \nonumber \\[1ex]
[\varepsilon^{12}_{mn}\varepsilon^{12}_{m'n'}] & = & 0  
\end{eqnarray}
where $l^{f}_{mn}$ and $l^{b}_{mn}$ are the mean free paths for 
forward and backward scattering from channel $m$ into channel $n$
and $1/l^{f/b}=\sum_{n} 1/l^{f/b}_{mn}$ are the total inverse mean 
free paths. The inverse mean free paths $1/l^{f}_{mn}$ and $1/l^{b}_{
mn}$ are defined by the probabilities per length $[|t_{mn}-\delta_{mn}
|^{2}]$ and $[|r_{mn}|^{2}]$ for a forward and a backward scattering 
process respectively. By Eq. (\ref{DefT}) and Eq. (\ref{Texpansion}) $[
|t_{mn}-\delta_{mn}|^{2}]=[|\varepsilon^{11}_{mn}|^{2}]$ and $[|r_{mn}
|^{2}]=[|\varepsilon^{12}_{mn}|^{2}]$ which leads to the above 
identification of the model parameters with the mean free paths.

Now we calculate $[\varepsilon^{ij}_{mn}]$ and $[\varepsilon^{ij}_{mn
}\varepsilon^{i'j'}_{m'n'}]$ for the Hamiltonian model (\ref{Mod1b}).  
Inserting $\bbox{\gamma}^{11}(x_{0})$ and $\bbox{\gamma}^{12}(x_{0})$ 
from Eq. (\ref{Tfirstorder}) into Eq. (\ref{T2expansion}) and averaging 
over the Gaussian white noise yields 
\begin{eqnarray}
\label{MomMod1b}
[\varepsilon^{ij}_{mn}] & = & 0 \nonumber\\[1ex]
[\varepsilon^{ij}_{mn}\varepsilon^{i'j'}_{m'n'}] & = & 
\frac{c(ij,i'j')}{l_{mn}}\frac{\delta_{mm'}\delta_{nn'}+\delta_{mn'}
\delta_{nm'}}{1+\delta_{mn}}
\end{eqnarray} 
where $1/l_{mn}\equiv 1/l^{f}_{mn}=1/l^{b}_{mn}=(m_{0}/(\hbar^{2}k))^{2} 
U_{mn}(1+\delta_{mn})$ and $1/l\equiv \sum_{m} 1/l_{mn}$. Note that 
$l^{f}_{mn}$ and $l^{b}_{mn}$ are not independent as in Eq. 
(\ref{MomMod2special}). The coefficients $c(ij.i'j')$ which are not 
related through the symmetries of $\bbox{\gamma^{ij}}(x_{0})$ are 
$c(11,11)=-c(11,22)=-c(12,21)=-1$, $c(11,21)=-c(11,12)^{*}=\exp(i2kx_{0}
)$, $c(12,12)=-\exp(-i4kx_{0})$. Hence, the moments 
$[\varepsilon^{ij}_{mn}\varepsilon^{i'j'}_{m'n'}]$ which vanish in model 
(\ref{MomMod2special}) oscillate with $x_{0}$ in Eq. (\ref{MomMod1b}). 
This will cause the coefficients of the FP equation (\ref{FPgl1}) for the 
probability distribution of $\bbox{TT}^{\dagger}$ to oscillate with the 
system length. In the limit of weak disorder $(kl\gg 1)$ these 
oscillations are very fast on the scale of the mean 
free path $l$ which is the characteristic length over which the 
probability distribution changes. Then, it is justified to average over 
the oscillations, which amounts to replace the oscillating moments in 
Eq. (\ref{MomMod1b}) by zero. The resulting model is very similar to the
model (\ref{MomMod2special}) if $l^{f}_{mn}=l^{b}_{mn}$ but not 
equivalent. It would be equivalent if the phases of $\eta_{mn}$ were
not random but had been chosen to take the values $\exp(i\pi/2)$ and 
$\exp(-i\pi/2)$ with equal probability. For stronger disorder it is no
longer justified to average over the oscillations.

As an alternative to the continuum limit of Mello and Tomsovic
one may specify directly the statistics of $\gamma^{ij}_{mn}(x)$.
Choosing Gaussian white noise such that
\begin{eqnarray}
\label{Mod3}
\langle \gamma^{ij}_{mn}(x) \rangle & = & 0 \nonumber \\  
\langle \gamma^{ij}_{mn}(x)\gamma^{i'j'}_{m'n'}(x') \rangle & = & 
\sigma^{ij,i'j'}_{mn,m'n'}(x)\delta(x-x') 
\end{eqnarray}
leads to
\begin{eqnarray}
\label{MomMod3}
[\varepsilon^{ij}_{mn}(x_{0})] & = & \sigma^{ik,kj}_{ml,ln}(x_{0})/2 
\nonumber\\[1ex] 
[\varepsilon^{ij}_{mn}(x_{0})\varepsilon^{i'j'}_{m'n'}(x_{0})] & = & 
\sigma^{ij,i'j'}_{mn,m'n'}(x_{0}).
\end{eqnarray}
The Hamiltonian model (\ref{Mod1b}) and the model (\ref{MomMod2special})
are special cases of this class of models. Note however that 
$[\varepsilon^{ij}_{mn}(x_{0})]$ differs in general from the result 
(\ref{MomMod2}) of the continuum limit. 
\section{Langevin and Fokker-Planck equations}
\label{SLvFPgl}
The evolution of the transfer matrix with the system length is a 
stochastic process which can be described by Langevin and FP equations. 
Dorokhov \cite{Dorokhov83a} recognized that the stochastic equations for
the matrix 
\begin{eqnarray}
\bbox{M} & = & \bbox{TT}^\dagger \nonumber \\[1.5ex]
& = & \left(\begin{array}{cc} \bbox{u} & \bbox{0} \\
                  \bbox{0} & \bbox{u}^*  \end{array}\right)
      \left(\begin{array}{cc} 
            \cosh\bbox{\Gamma} & \sinh\bbox{\Gamma} \\
            \sinh\bbox{\Gamma} & \cosh\bbox{\Gamma}  
            \end{array}\right)
      \left(\begin{array}{cc} \bbox{u}^\dagger & \bbox{0} \\
                  \bbox{0} & \bbox{u}^T  \end{array}\right) ,
\end{eqnarray}
are closed which allows to eliminate the degrees of freedom of 
$\bbox{v}$. We follow Dorokhov and start from his Langevin equations 
(\ref{Lvgl2}) for $\bbox{\Gamma}$ and $\bbox{u}$ which are derived in 
Appendix \ref{ALvgl} for the sake of completeness. Due to the 
self-averaging of the $\alpha_{m}$ one expects that in the insulating 
regime where $L\gg\xi=1/(2\alpha^{\infty}_{1})$ the $\Gamma_{m}$ can be 
ordered: $1\ll\Gamma_{1}\ll\cdots\ll\Gamma_{N}$. This ordering justifies
the neglect of exponentially small contributions to the terms
\begin{eqnarray}
\label{approx}
\coth\Gamma_j & = & 1+O(\exp(-2\Gamma_{j}))  \nonumber \\[1.5ex]
\frac{\sinh\Gamma_n}{\cosh\Gamma_n-\cosh\Gamma_j}  & = &
\left\{ \begin{array}{ccc} 
         0+O(\exp(\Gamma_{n}-\Gamma_{j})) &  & n<j \\
         1+O(\exp(\Gamma_{j}-\Gamma_{n})) &  & n>j 
         \end{array}\right.\hspace{0.5cm} ,
\end{eqnarray}
of the general Langevin equations (\ref{Lvgl2}) which leads to the 
considerable simplification 
\begin{eqnarray}
\label{Lvgl1}
\frac{d\Gamma_{m}}{dL} & = & E_{mm}+E^{*}_{mm} \nonumber \\[1.5ex]
\frac{d \bbox{u}}{dL} & = & \bbox{\gamma}^{11}\bbox{u}+\bbox{u}\bbox{P}. 
\end{eqnarray}
The matrix elements of $\bbox{P}$ are $P_{mn}=\theta (n-m) E_{mn}-
\theta (m-n) E^*_{mn}$ where
\begin{equation}
\theta (n) =  
\left\{\begin{array}{ccc} 1           &  & n>0 \\
                          \frac{1}{2} &  & n=0 \\
                          0           &  & n<0
        \end{array} \right. \hspace{0.5cm} 
\end{equation}
and $\bbox{E}=\bbox{u}^{\dagger}\bbox{\gamma}^{12}\bbox{u}^{*}$. The 
symmetries $\bbox{\gamma}^{\dagger}=-\bbox{\gamma}$ and $\bbox{P}^{
\dagger}=-\bbox{P}$ imply that $d(\bbox{uu}^{\dagger})/dL|_{\bbox{uu}^{
\dagger}=\bbox{1}}=0$. Thus, the simplified Langevin 
equations (\ref{Lvgl1}) still conserve the unitarity of $\bbox{u}$. The 
stochastic process described by them leads to a limiting distribution 
for $(\Gamma_{m}-2\alpha^{\infty}_{m}L)/\sqrt{L}$ which is 
independent of the initial conditions. Hence, this limiting 
distribution must be identical to the one which is produced by the 
original Langevin equations (\ref{Lvgl2}) as long as $1\ll\Gamma_{1}\ll
\cdots\ll\Gamma_{N}$. Therefore, it is possible to use the simplified 
Langevin equations together with convenient initial conditions to 
determine the form of the limiting distribution in this parameter range.

Due to the neglect of exponential small terms in (\ref{approx}), 
$\Gamma_{m}$ can become negative and it is natural to extend the range 
of $\Gamma_{m}$ to $-\infty$ which is justified because the probability 
to find a negative value of $\Gamma_{m}$ will turn out to be 
exponentially small. For similar reasons we relax the strict ordering of
the $\Gamma_{m}$. A parametrization of $\bbox{u}$ by a set of $N^{2}$ 
independent parameters seems to be rather complicated. Instead, we 
extend the range of the matrix elements $u_{mn}=x_{mn}+i y_{mn}$ to 
arbitrary complex numbers thereby obtaining a stochastic process on a 
higher dimensional cartesian space. The standard derivation technique 
\cite{Risken89} for the FP equation of such a process yields   
\begin{equation}
\label{FPgl1}
\frac{\partial p(L;\bbox{\Gamma},\bbox{u},\bbox{u}^*)}{\partial L}  =
(\partial_{\Gamma_m}\partial_{\Gamma_n}\hat{A}_{mn}+\partial_{\Gamma_m}
\hat{B}_m + \hat{C}) p(L;\bbox{\Gamma},\bbox{u},\bbox{u}^*) ,
\end{equation}
where $\prod_{m}\mbox{d}\Gamma_{m}\prod_{m'n'}\mbox{d}x_{m'n'}\mbox{d}
y_{m'n'}$ is the measure of the cartesian space. The operators 
$\hat{A}_{mn}$, $\hat{B}_m$, and $\hat{C}$ are
\begin{eqnarray}
\label{FPOp1}
\hat{A}_{mn} & = & \frac{1}{2} [\Delta\Gamma_m\Delta\Gamma_n] 
\nonumber\\[1.5ex]
\hat{B}_{m} & = & -[\Delta\Gamma_{m}]+\partial_{u_{m'n'}}[\Delta
\Gamma_{m}\Delta u_{m'n'}]+\partial_{u^{*}_{m'n'}}[\Delta\Gamma_{m}
\Delta u^{*}_{m'n'}]\nonumber \\[1.5ex]
\hat{C} & = & -\partial_{u_{mn}}[\Delta u_{mn}]-\partial_{u_{mn}^*}
[\Delta u_{mn}^*] +\partial_{u_{mn}}\partial_{u_{m'n'}^*}
[\Delta u_{mn}\Delta u_{m'n'}^*] \nonumber \\ &  &
+\frac{1}{2}\partial_{u_{mn}}\partial_{u_{m'n'}}[\Delta u_{mn}\Delta 
u_{m'n'}]+\frac{1}{2}\partial_{u_{mn}^*}\partial_{u_{m'n'}^*}[\Delta 
u_{mn}^*\Delta u_{m'n'}^*]  ,
\end{eqnarray}
where $\Delta\Gamma_m=\Gamma (L+\delta L)-\Gamma (L)$ and  $\Delta 
u_{mn}=u_{mn}(L+\delta L)-u_{mn}(L)$. The brackets $[\cdots]$ define the 
coefficients of the FP operator and stand for $\lim_{\delta L\to 0}
\langle\cdots\rangle_{\delta L}/\delta L$ where $\langle\cdots\rangle_{
\delta L}$ is the average over the disorder in the region between $L$ 
and $L+\delta L$. The explicit form of the coefficients is derived in 
appendix \ref{AFPkoeff}.
  
The multiplication of a probability distribution on the cartesian space 
by the $\delta$-function
\begin{eqnarray}
\label{deltafunction}
\delta(\bbox{1},\bbox{uu}^{\dagger}) & = &
\prod_{m=1}^{N}\delta\left(1-\sum_{n} u_{mn}u^{*}_{mn}\right)\prod_{
m'<n'}\delta\left(\sum_{n} \mbox{Re}(u_{m'n}u^{*}_{n'n})\right)  
\delta\left(\sum_{n} \mbox{Im}(u_{m'n}u^{*}_{n'n})\right)   \nonumber \\
&  &  
\end{eqnarray}
restricts it to the unitary group. Since the Langevin equations 
(\ref{Lvgl1}) conserve unitarity, one expects that the FP operator 
commutes with the $\delta$-function. In fact, the operators $\hat{A}$, 
$\hat{B}$ and $\hat{C}$ commute with every function of the type $f(
\bbox{uu}^{\dagger})$ for {\it arbitrary} complex matrices $\bbox{u}$. A
lengthy but straightforward calculation with the coefficients (
\ref{FPkoeff}) which exploits the symmetries (\ref{symm2}) proves that 
this is true. Thus, the restriction to the unitary group may be 
incorporated into the new measure $\prod_{m}\mbox{d}\Gamma_{m}d\mu(
\bbox{u})$ where $d\mu(\bbox{u})={\cal V}^{-1}(N)\delta(\bbox{1},\bbox{u
 u}^{\dagger})\prod_{m'n'}\mbox{d}x_{m'n'} \mbox{d}y_{m'n'}$ and ${\cal 
V}(N)$ is the volume of the unitary group. It is shown in appendix 
\ref{AInvMeas} that $d\mu(\bbox{u})$ is the invariant measure of $U(N)$.
We note that an integral of the type $\int\mbox{d}\mu(\bbox{u})\hat{X}g(
\bbox{u},\bbox{u}^{*}) $ ($\hat{X}=\hat{A}$, $\hat{B}$, or $\hat{C}$) 
with respect to the invariant measure is evaluated in two steps. First, 
the $\delta$-function is commuted with $\hat{X}$. Second, the 
integrations are carried out yielding only contributions for terms which
do not have derivatives with respect to $u_{mn}$ or $u_{mn}^{*}$ in 
front of them.

It is worth emphasizing that the operators $\hat{A}$, $\hat{B}$ and 
$\hat{C}$ do not depend on $\bbox{\Gamma}$. This is the great 
simplification which has been achieved by the neglect of the exponential
small terms in (\ref{approx}). However, they can still depend on the 
system length $L$ as it is the case for the Hamiltonian model 
(\ref{Mod1b}).  There, the factor $\exp(-ikx_{0})$ of $\bbox{\gamma}^{
12}(x_{0})$ in Eq. (\ref{Tfirstorder}) leads to terms which oscillate 
with the system length. For weak disorder ($kl\gg 1$) the oscillations 
are fast on the scale of the mean free path $l$ which justifies 
replacing the oscillating terms by their averages. For stronger disorder
the oscillations can be absorbed into the new variable $\tilde{u}_{mn}=
u_{mn}\exp(ikL)$. The transformation of the FP equation to this variable
leads to the additional term $-ik(\tilde{u}_{mn}\partial_{\tilde{u}_{mn}
}-\tilde{u}^{*}_{mn}\partial_{\tilde{u}^{*}_{mn}})$ in the FP operator. 
The resulting FP equation is very similar to the FP equation of Kree and
Schmid \cite{Kree81a}. It is the generalization from $N=1$ to arbitrary 
channel numbers. 
\section{The limiting distribution of the transfer matrix}
\label{Slimdistr}
Kree and Schmid discussed thoroughly the asymptotic probability 
distribution for the Landauer conductance  $g=|t|^{2}$ of a long 
one-dimensional wire \cite{Kree81a}. The transmission and reflection 
amplitude for incident waves from the right are $t=u(2/(1+\cosh\Gamma)
)^{1/2}v$ and $r=u^{2}((\cosh\Gamma-1)/(\cosh\Gamma+1))^{1/2}$ where $u$
and $v$ are simply phases. They showed that $(\Gamma-2\alpha^{\infty}L)/
\sqrt{L}$ has a Gaussian limiting distribution if $L\to\infty$ which 
implies a log-normal distribution for the conductance. The parameters of
the Gaussian distribution were expressed in terms of averages over the 
stationary distribution of $u$. Similarly we expect that the 
corresponding quantity $(\bbox{\Gamma}-2\bbox{\alpha}^{\infty}L)/
\sqrt{L}$ for the quasi-one dimensional wire has a multivariate 
Gaussian limiting distribution whose parameters can be expressed in 
terms of averages over the stationary distribution of $\bbox{u}$.

It is useful to look at the first moments $\langle\Gamma_{m}\rangle_{L}$
in some detail before deriving the general form of the limiting 
distribution. Integrating Eq. (\ref{FPgl1}) with respect to $\bbox{
\Gamma}$ leads to the closed FP equation
\begin{equation}
\frac{\partial q(L;\bbox{u},\bbox{u}^{*})}{\partial L}=\hat{C}
q(L;\bbox{u},\bbox{u}^{*})
\end{equation}
for $\bbox{u}$. We expect that the stationary distribution of $\bbox{u}$
is the unique stationary solution $q_{stat}(\bbox{u},\bbox{u}^{*})$ of 
this equation.
Hence, the spectrum of $\hat{C}$ should consist of one eigenvalue $\nu_{
0}$ which is zero and others  $\nu_{i}$ with negative real parts. The 
smallest absolute value of these real parts is henceforth called $\nu$. 
In the sequel it will be assumed that the eigenfunctions $q_{i}(\bbox{u}
,\bbox{u}^{*})$ of $\hat{C}$ form a complete system where $q_{0}(\bbox{u
},\bbox{u}^{*})=q_{stat}(\bbox{u},\bbox{u}^{*})$. Then, any probability 
distribution may be expanded into $\sum_{i}c_{i}q_{i}(\bbox{u},\bbox{u
}^{*})$ where conservation of probability implies $\int d\mu(\bbox{u}) 
q_{i}(\bbox{u},\bbox{u}^{*})=0$ for $i\neq 0$ and $c_{0}=1$. Multiplying
Eq. (\ref{FPgl1}) with $\Gamma_{m}$ and integration by parts yields 
\begin{eqnarray}
\label{firstMom}
\frac{\partial\langle\Gamma_{m}\rangle_L}{\partial L} & = &
-\int d\mu (\bbox{u}) \hat{B}_{m} q(L;\bbox{u},\bbox{u}^*) \nonumber \\
& = & \int d\mu (\bbox{u}) [\Delta\Gamma_{m}] q(L;\bbox{u},\bbox{u}^*) . 
\end{eqnarray}
Expanding the initial distribution into eigenfunctions $q(0;\bbox{u},
\bbox{u}^{*})=q_{stat}(\bbox{u},\bbox{u}^{*})+\sum_{i\neq 0}c_{i}$ $q_{i
}(\bbox{u},\bbox{u}^{*})$ gives $q(L;\bbox{u},\bbox{u}^{*})=q_{stat}( 
\bbox{u},\bbox{u}^{*})+\sum_{i\neq 0} c_{i}\exp(\nu_{i}L)q_{i}(\bbox{u},
\bbox{u}^{*})$. This leads to the large length asymptotic behaviour
\begin{equation}
\label{firstMoment}
\langle\Gamma_{m}\rangle_{L}-\langle\Gamma_{m}\rangle_{0}\approx 
L\int d\mu(\bbox{u}) [\Delta\Gamma_{m}] q_{stat}(\bbox{u},\bbox{u}^*)+
\left[const.+O(\exp\{-\nu L\})\right] 
\end{equation}
where the terms in brackets result from the crossover of the initial
into the stationary distribution of $\bbox{u}$. The self-averaging 
$\alpha^\infty_{m}\equiv\lim_{L\to\infty}\alpha_{m}(L)=
\lim_{L\to\infty}\langle\alpha_{m}(L)\rangle_{L}$ of the Lyapunov 
exponents implies $\alpha^\infty_{m}=\lim_{L\to\infty}\langle\Gamma_{m}
\rangle_{L}/2L$ which leads to
\begin{equation}
\label{LyaExp}
\alpha^{\infty}_{m}=\frac{1}{2}\int d\mu(\bbox{u}) 
[\Delta\Gamma_{m}] q_{stat}(\bbox{u},\bbox{u}^*).  
\end{equation}
This relation has been first derived by Dorokhov \cite{Dorokhov83a}. 
Using Eq. (\ref{FPkoeff}) to calculate $[\Delta\Gamma_{m}]$ for the 
model (\ref{MomMod2special}) yields  
\begin{equation} 
\alpha^{\infty}_{m} = 
\frac{\theta(m-k_3)}{2l^{b}_{k_1k_2}(1+\delta_{k_1k_2})}\int d\mu(
\bbox{u})
\begin{array}[t]{l}
\left(u_{k_1m}u_{k_2k_3}u^{*}_{k_1k_3}u^{*}_{k_2m}+u^{*}_{k_1m}u^{*}_{
k_2k_3} u_{k_1k_3}u_{k_2m} \right . \\
\left. +2 u_{k_1m}u_{k_2k_3}u^{*}_{k_1m}u^{*}_{k_2k_3}\right)
q_{stat}(\bbox{u},\bbox{u}^{*}).\end{array}
\end{equation}
The same formula has been obtained by Chalker and Bernhardt 
\cite{Chalker93a} for the special case that there is only back 
scattering into the same channel. They discussed also the consequences 
of this relation in the context of the Anderson transition.

In the sequel we will go beyond the first moment $\langle\Gamma_{m} 
\rangle_{L}$ and derive the general form of the limiting distribution of 
$(\bbox{\Gamma}-2\bbox{\alpha}^{\infty}L)/\sqrt{L}$. For the sake of 
simplicity we choose the initial distribution
\begin{equation}
p(0;\bbox{\Gamma},\bbox{u},\bbox{u}^{*})=\prod_{m=1}^{N}\delta(\Gamma_{m
})q_{stat}(\bbox{u},\bbox{u}^{*})
\end{equation} 
which implies that $\langle\Gamma_{m}\rangle_{L}=2\alpha_{m}^{\infty}L$ 
(see Eq. (\ref{firstMoment})). The formalism which will be devlopped 
below could be used to show that different initial condition would not 
change the form of the limiting distribution but only the way it is 
approached. It is convenient to introduce Dirac notation
\begin{eqnarray}
(0|\hat{C}=0 &\hspace{1.5cm} & \hat{C}|0)=0 \nonumber \\
\hat{\cal P}=|0)(0| &\hspace{1.5cm}& \hat{\cal Q}=1-\hat{\cal P} 
\end{eqnarray}
where $q_{stat}(\bbox{u},\bbox{u}^{*})=(\bbox{u},\bbox{u}^{*}|0)$ and
$(0|\bbox{u},\bbox{u}^{*})=(0|0)=1$ so that the average $\int d\mu(
\bbox{u})$ $\hat{X}q_{stat}(\bbox{u},\bbox{u}^{*})$ may be simply 
expressed as $(0|\hat{X}|0)$.
 
The central quantity which will be used below to derive the limiting 
distribution is the generating function 
\begin{equation}
\label{Genfunc1}
P(L;\bbox{\tau},\bbox{u},\bbox{u}^*)=\int\prod_{m'} d\Gamma_{m'} 
\exp\{i\sum_m \tau_m(\Gamma_{m}-\langle\Gamma_{m}\rangle_{L})\} 
p(L;\bbox{\Gamma},\bbox{u},\bbox{u}^*)
\end{equation}
for the central moments 
\begin{eqnarray}
\label{Moments1}
{\cal C}_{L}(m_{1}r_{1},\cdots,m_{k}r_{k}) & = &
\langle(\Gamma_{m_{1}}-\langle\Gamma_{m_{1}}\rangle_{L})^{r_{1}}\cdots
(\Gamma_{m_{k}}-\langle\Gamma_{m_{k}}\rangle_{L})^{r_{k}}\rangle_{L}
\nonumber\\
& = & \int\mbox{d}\mu(\bbox{u})(-i\partial_{\tau_{m_1}})^{r_1}\cdots 
(-i\partial_{\tau_{m_k}})^{r_k} P(L;\bbox{\tau},\bbox{u},\bbox{u}^*)
\left.\right|_{\bbox{\tau}=0}.
\end{eqnarray}
The Fourier transform of the generating function gives back the 
probability distribution
\begin{equation}
\label{FourTr1}
p(L;\bbox{\Gamma},\bbox{u},\bbox{u}^{*}) = \frac{1}{(2\pi)^{N}}\int 
\prod_{m=1}^{N} d\tau_{m} \exp\left\{-i\sum_{n=1}^{N}(\Gamma_{n}-
\langle\Gamma_{n}\rangle_{L})\tau_{n}\right\} 
P(L;\bbox{\tau},\bbox{u},\bbox{u}^*). 
\end{equation}
This implies the evolution equation 
\begin{equation}
\label{EvoGF}
\frac{\partial P(L;\bbox{\tau},\bbox{u},\bbox{u}^*)}{\partial L}=
(-\tau_m\tau_n \hat{A}_{mn}-i\tau_m\hat{B}^{0}_m+\hat{C})P(L;\bbox{\tau}
,\bbox{u},\bbox{u}^*)                                       
\end{equation}
where  
\begin{eqnarray}
\label{DefB0m}
\hat{B}^{0}_{m} & = & \hat{B}_{m}+\frac{d\langle\Gamma_{m}\rangle_{L}}{
dL} \nonumber \\
& = & \hat{B}_{m}-(0|\hat{B}_{m}|0)
\end{eqnarray}
and $P(0;\bbox{\tau},\bbox{u},\bbox{u}^*)=q_{stat}(\bbox{u},\bbox{u}^{*}
)$. The formal solution of Eq. (\ref{EvoGF}) is
\begin{equation}
P(L;\bbox{\tau},\bbox{u},\bbox{u})=\exp\{(-\tau_m\tau_n\hat{A}_{mn}-i
\tau_m\hat{B}^{0}_m+\hat{C})L\}q_{stat}(\bbox{u},\bbox{u}^{*}).
\end{equation}
We follow Kree and Schmid and  reexpress it in terms of the operator 
generalization of the Cauchy formula  $\exp (bL)=1/(2\pi i)\oint 
d\zeta\exp(i\zeta L)/(\zeta+ib)$, 
\begin{equation}
P(L;\bbox{\tau},\bbox{u},\bbox{u}^{*})=
\frac{1}{2\pi i}\oint d\zeta\exp\{i\zeta L\}\hat{R}(\bbox{\tau},\zeta)
q_{stat}(\bbox{u},\bbox{u}^{*})
\end{equation}
where
\begin{equation}
\hat{R}(\bbox{\tau},\zeta)=\left(\zeta+i(-\tau_{m}\tau_{n}\hat{A}_{mn}-
i\tau_{m}\hat{B}^{0}_{m}+\hat{C})\right)^{-1}
\end{equation}
is the resolvent operator and the integration contour encircles all
the eigenvalues of $-i(-\tau_{m}\tau_{n}\hat{A}_{mn}-i\tau_{m}\hat{B}^{0
}_{m}+\hat{C})$ counterclockwise. Equation (\ref{Moments1}) for the 
moments of $(\Gamma_{m}-\langle\Gamma_{m}\rangle_{L})$ only requires the
resolvent operator for infinitesimal values of $\tau$. The spectrum of $
-i(-\tau_{m}\tau_{n}\hat{A}_{mn}-i\tau_{m}\hat{B}^{0}_{m}+\hat{C})$ then
lies in the neighbourhood of the spectrum of $-i\hat{C}$ which consists 
of one zero eigenvalue  and other eigenvalues in the upper half plane. 
Hence one can choose the integration to run just below the real line 
from minus to plus infinity and close the contour in the upper half 
plane.

We do not reconstruct the limiting distribution  from the moments 
${\cal C}_{L}(m_{1}r_{1},\cdots,m_{k}r_{k})$ but proceed indirectly. 
First consider the linear combination $\Omega\equiv c_{m} \Omega_{m}
\equiv c_{m}(\Gamma_{m}-\langle\Gamma_{m}\rangle_{L})$ with arbitrary 
coefficients $c_{m}$ and study its moments which are given by 
\begin{eqnarray}
\langle\Omega^{r}\rangle & = & \int d\mu(\bbox{u})(-i\partial_{\tau}
)^{r}P(L;\tau\bbox{c},\bbox{u},\bbox{u}^{*})|_{\tau=0}\nonumber\\ 
& = & \frac{1}{2\pi i} \oint d\zeta\exp\{i\zeta L\}(-i\partial_{\tau}
)^{r} (0|\hat{R}(\tau\bbox{c},\zeta)|0)|_{\tau=0}.
\end{eqnarray}
Since 
\begin{equation}
\hat{R}(\tau\bbox{c},\zeta)=
\left(\zeta+i(-\tau^{2} c_{m}c_{n}\hat{A}_{mn}-i\tau c_{m}\hat{B}^{0
}_{m}+\hat{C}\right)^{-1}
\end{equation}
is similar to the resolvent operator of a one-dimensional wire one can 
employ the technique of Kree and Schmid to calculate the moments. 
Expanding $\hat{R}(\tau\bbox{c},\zeta)$ into powers of $\tau$ yields 
\begin{equation}
\label{ResExp}
\hat{R}(\tau\bbox{c},\zeta)=\hat{R}_{0}(\zeta)\sum_{k=0}^{\infty}\left[
(i\tau^{2}
c_{m}c_{n}\hat{A}_{mn}-\tau c_{m}\hat{B}^{0}_{m})\hat{R}_{0}(\zeta)
\right]^{k}
\end{equation} 
where $\hat{R}_{0}(\zeta)=(\zeta+i\hat{C})^{-1}$ may be decomposed into 
a part which is singular at $\zeta=0$ and a non-singular part
\begin{eqnarray}
\hat{R}_{0} & = & \hat{R}_{0s}+\hat{R}_{0n} \nonumber\\
& = & \frac{\hat{\cal P}}{\zeta}+\frac{\hat{\cal Q}}{\zeta+i\hat{C}}.
\end{eqnarray}
Only the terms of order $\tau^{r}$ of the expansion (\ref{ResExp}) 
contribute to the $r^{th}$ moment of $\Omega$. The first moment  
\begin{eqnarray}
\langle\Omega\rangle_{L} & = & 
\frac{1}{2\pi i} \oint d\zeta\exp\{i\zeta L\}(0|\hat{R}_{0}(\zeta) 
i c_{m}B_{m}^{0} \hat{R}_{0}(\zeta)|0)\nonumber\\ 
& = & \frac{c_{m}}{2\pi}(0|B_{m}^{0}|0)\oint d\zeta\frac{\exp\{i\zeta L
\}}{\zeta^{2}}
\end{eqnarray}
is zero because of Eq. (\ref{DefB0m}) as it should be. Collecting 
the terms which contribute to the second moment yields
\begin{eqnarray} 
\langle\Omega^{2}\rangle  & = &
\frac{1}{2\pi i} \oint d\zeta\exp\{i\zeta L\}
(-2c_{m}c_{n})(0|i\hat{R}_{0}A_{mn}\hat{R}_{0}
+\hat{R}_{0}B_{m}^{0}\hat{R}_{0}B^{0}_{n}\hat{R}_{0}|0)
\nonumber\\
& = & \frac{1}{2\pi i} \oint d\zeta\exp\{i\zeta L\}
\frac{-2 c_{m}c_{n}}{\zeta^{2}}(0|i A_{mn}
+B_{m}^{0}\frac{\hat{\cal Q}}{\zeta+i\hat{C}}
B^{0}_{n}|0)
\end{eqnarray}
The residues of the pole of second order at $\zeta=0$ are $2 c_{m}c_{n}
(0|\hat{A}_{mn}|0) L$ and $-2 c_{m}c_{n}$ $(0|\hat{B}^{0}_{m}\hat{C}^{-1
}\hat{B}^{0}_{n} L+\hat{B}^{0}_{m}\hat{C}^{-2}\hat{B}^{0}_{n}|0)$
for the first and the second term respectively. The poles in the upper
half plane of the second term only give rise to exponential decaying 
contributions. Hence 
\begin{equation}
\label{varOmega}
\langle\Omega^{2}\rangle = 2 \omega L + const. + 
O\left(\exp\{-\nu L\}\right)
\end{equation}
where
\begin{eqnarray}
\omega & = & c_{m}c_{n} (0|\hat{A}_{mn}-
\hat{B}^{0}_{m}\hat{C}^{-1}\hat{B}^{0}_{n}|0)\nonumber\\
& = & c_{m}c_{n}{\cal A}_{mn}.
\end{eqnarray}
It can be shown along the same lines as in work of Kree and Schmid that 
the higher moments have the form
\begin{eqnarray}
\label{Moments2}
\langle\Omega^{2n}\rangle & = & \frac{(2n)!}{n!}(\omega L)^{n}+
O(L^{n-1}) \nonumber \\
\langle\Omega^{2n+1}\rangle & = & O(L^{n}).
\end{eqnarray}
For details of the proof we refer to \cite{Kree81a}. The form of the
moments implies that $\bar{\Omega}\equiv\Omega/\sqrt{L}$ has the 
Gaussian limiting  distribution  
\begin{equation}
\label{pOmega}
s_{\infty}(\bar{\Omega})=\frac{1}{\sqrt{4\pi\omega}} 
\exp\left\{-\bar{\Omega}^{2}/(4\omega)\right\}. 
\end{equation}
Now we discard the finite length corrections to the limiting 
distribution and keep only the universal parts $\langle\Omega^{2n}
\rangle^{u}_{L}=(\omega L)^{n}(2n)!/n!$ and $\langle\Omega^{2n+1}
\rangle^{u}_{L}=0$ of the moments (\ref{Moments2}). The corresponding 
universal part of the generating function $S(L;\bbox{\tau})=\int d\mu(
\bbox{u})P(L;\bbox{\tau},\bbox{u},\bbox{u}^{*})$ is denoted by $S^{u}(L;
\bbox{\tau})$. Since $\langle\Omega^{r}\rangle^{u}_{L}=(-i\partial
_{\tau})^{r}S^{u}(L;\tau\bbox{c})|_{\tau=0}$ we find
$S^{u}(L;\tau\bbox{c})=\exp\{-\tau^{2}c_{m}c_{n}{\cal A}_{mn}L\}$ which
implies
\begin{equation}
S^{u}(L;\bbox{\tau})=\exp\{-\tau_{m}\tau_{n}{\cal A}_{mn}L\}.
\end{equation} 
The Fourier transform (\ref{FourTr1}) of $S^{u}(L;\bbox{\tau})$ yields
the universal part of the probability distribution  
\begin{equation}
\label{limdistr1}
s^{u}(L;\bbox{\Gamma})=\frac{1}{(4\pi L)^{N/2}\sqrt{\det(\{{\cal A}_{mn}
\})}}\exp\left\{-(\Gamma_{m}-\langle\Gamma_{m}\rangle_{L}){\cal 
A}^{-1}_{mn}(\Gamma_{n}-\langle\Gamma_{n}\rangle_{L})/4L\right\}
\end{equation}
where $\langle\Gamma_{m}\rangle_{L}=2\alpha^{\infty}L$. Hence, the 
limiting distribution of $\bar{\bbox{\Omega}}\equiv(\bbox{\Gamma}-2
\bbox{\alpha}^{\infty}L)/\sqrt{L}$ is 
\begin{equation}
s_{\infty}(\bar{\bbox{\Omega}})=\frac{1}{(4\pi)^{N/2}\sqrt{\det(\{{\cal
A}_{mn}\})}} \exp\left\{ -\bar{\Omega}_{m}{\cal A}^{-1}_{mn}\bar{\Omega
}_{n}/4\right\}.
\end{equation}
Note, that the form of the correlator
\begin{eqnarray}
\langle (\alpha_{m}-\alpha^{\infty}_{m})(\alpha_{n}-\alpha^{\infty}_{n}) 
\rangle^{u}_{L} 
& = & \frac{1}{4L^{2}}(-i\partial_{\tau_{m}})(-i\partial_{\tau_{n}})
S^{u}(L;\bbox{\tau})|_{\tau=0}\nonumber\\
& = & \frac{1}{4L}({\cal A}_{mn}+{\cal A}_{nm})
\end{eqnarray}
implies that the $\alpha_{m}\equiv\Gamma_{m}/2L$ are self-averaging and 
that the fluctuations around their limiting values are in general 
correlated. Such correlations are not predicted by RMT or the DMPK 
equation but are consistent with numerical simulations 
\cite{Markos93a,Markos94a}.

The variance of the Gaussian distribution of $\ln g\approx -\Gamma_{1}+ 
const.$ in the insulating regime which follows from $\mbox{var}(\ln 
g)\approx\mbox{var}(\Gamma_{1})$ and Eq. (\ref{varOmega}) is
\begin{equation}
\mbox{var}(\ln g)\approx 2{\cal A}_{11}L+O(1). 
\end{equation}
\section{The Equivalent Channel Model as a special case}
\label{SECM}
Mello and Tomosovic have shown \cite{Mello91c,Mello92a} that the 
joint probability distribution $s(L;\bbox{\Gamma})$ of the ECM which is 
the model (\ref{MomMod2special}) with backscattering mean free paths of 
the form
\begin{equation}
\label{mfpECM}
\frac{1}{l^{b}_{mn}}=\frac{1+\delta_{mn}}{l^{b}(N+1)}
\end{equation}
obeys the DMPK equation for $\beta=1$. The form (\ref{RMTGauss}) of the
solution for the DMPK 
equation in the insulating regime is a special case of the multivariate 
Gaussian distribution (\ref{limdistr1}). Hence, we expect to recover 
this solution from our approach if we apply it to the ECM.

Evaluating Eq. (\ref{FPkoeff}) for the coefficients $[\Delta\Gamma_{m}
\Delta\Gamma_{n}]$ and $[\Delta\Gamma_{m}]$ with the backscattering
mean free paths (\ref{mfpECM}) yields
\begin{eqnarray}
[\Delta\Gamma_{m}\Delta\Gamma_{n}] & = & \frac{2}{l^{b}(N+1)} 
\left(u^{*}_{k_{1}m}u^{*}_{k_{2}m}u_{k_{1}n}u_{k_{2}n}
      +u_{k_{1}m}u_{k_{2}m}u^{*}_{k_{1}n}u^{*}_{k_{2}n}\right) 
\nonumber\\
\left[\Delta\Gamma_{m}\right] & = & 
\frac{\theta(m-k_{3})}{l^{b}(N+1)}
\begin{array}[t]{l}
\left(  u^{*}_{k_{1}m}u^{*}_{k_{2}k_{3}}u_{k_{1}k_{3}}u_{k_{2}m}
        +u_{k_{1}m}u_{k_{2}k_{3}}u^{*}_{k_{1}k_{3}}u^{*}_{k_{2}m} 
\right. \\
\left.  +u^{*}_{k_{1}m}u^{*}_{k_{2}k_{3}}u_{k_{2}k_{3}}u_{k_{1}m}
        +u_{k_{1}m}u_{k_{2}k_{3}}u^{*}_{k_{2}k_{3}}u^{*}_{k_{1}m} 
\right).  
\end{array}
\end{eqnarray} 
Since the coefficients have the constant values $[\Delta\Gamma_{m}
\Delta\Gamma_{n}]=4\delta_{mn}/(l^{b}(N+1))$ and $[\Delta\Gamma_{m}]=
2m/(l^{b}(N+1))$ if $\bbox{u}$ is unitary, one can integrate the 
evolution equation (\ref{EvoGF}) for the generating function $P(L;
\bbox{\tau},\bbox{u},\bbox{u}^{*})$ with respect to $\bbox{u}$. The 
solution of the resulting equation
\begin{equation}
\frac{\partial S(L;\bbox{\tau})}{\partial L}=-\frac{2\delta_{mn}}{l^{b}
(N+1)}\tau_{m}\tau_{n} S(L;\bbox{\tau})
\end{equation}
is $S(L;\bbox{\tau})=\exp\{-2\sum_{m}\tau_{m}^{2}L/(l^{b}(N+1))\}$ which
leads to the Gaussian distribution  
\begin{equation}
s(L;\bbox{\Gamma}) =  
\prod_{m=1}^{N}\frac{1}{\sqrt{2\pi\sigma^{2}}}\exp\left\{-\frac{(
\Gamma_{m}-\langle\Gamma_{m}\rangle_{L})^{2}}{2\sigma^{2}}\right\} 
\end{equation} 
with $\langle\Gamma_{m}\rangle_{L}=2mL/(l^{b}(N+1))$ and $\sigma^{2}=4L/
(l^{b}(N+1))$ as expected.
  
It is a specific property of ECMs that the evolution of the joint 
distribution $s(L;\bbox{\Gamma})$ decouples from $\bbox{u}$. Therefore, 
the limiting distribution does not depend on the stationary distribution 
of $\bbox{u}$. Still, it is of interest to know the stationary 
distribution. Solving the equation $\hat{C}q_{stat}(\bbox{u},\bbox{u}^{*
})=0$ which determines it is in general a difficult problem. We 
demonstrate below that due to the simple form of the backscattering mean
free paths the stationary probability measure $q_{stat}(\bbox{u},\bbox{u
}^{*})\prod_{m,n}dRe(u_{mn})dIm(u_{mn})$ can be found to be ${\cal V}^{
-1}\delta(\bbox{1},\bbox{uu}^{\dagger})\prod_{m,n}dRe(u_{mn})dIm(u_{mn})
$ which is the invariant measure of the unitary group.\\
Using the form (\ref{FPOp2}) of $\hat{C}$ in which the derivatives act 
directly on the distribution and the property that $\hat{C}$ commutes
with the $\delta$-function yields
\begin{eqnarray}
\label{partC4}
\hat{C}\delta(\bbox{1},\bbox{uu}^{\dagger}) & = & 
\delta(\bbox{1},\bbox{uu}^{\dagger}) 
\left(\frac{1}{2}mn[\Delta\Gamma_{m}\Delta\Gamma_{n}]-m[\Delta
\Gamma_{m}]\right)\nonumber \\  
& = & \delta(\bbox{1},\bbox{uu}^{\dagger}) g(\bbox{uu}^{\dagger})
\end{eqnarray}
where $g(\bbox{uu}^{\dagger})$ is zero if $\bbox{u}$ is unitary. 
Applying higher powers of $\hat{C}$ to the 
$\delta$-function gives $\hat{C}^{k}\delta(\bbox{1},\bbox{uu}^{\dagger})
=\delta (\bbox{1},\bbox{uu}^{\dagger})g(\bbox{uu}^{\dagger})^{k}$. 
Hence, the initial distribution $q(0;\bbox{u},\bbox{u}^{*})={\cal V}^{-1
}\delta(\bbox{1},\bbox{uu}^{\dagger})$ evolves into
\begin{eqnarray}
q(L;\bbox{u},\bbox{u}^{*}) & = & {\cal V}^{-1}
\delta(\bbox{1},\bbox{uu}^{\dagger})\exp\{g(\bbox{uu}^{\dagger})L\}
\nonumber\\
& = & {\cal V}^{-1}\delta(\bbox{1},\bbox{uu}^{\dagger})
\end{eqnarray}
which shows that it is the stationary distribution.
\section{Conclusion}
The general form (\ref{limdistr1}) of the limiting distribution and the 
link between its parameters and the stationary distribution of $\bbox{u}
$ are the main results of this paper. This form implies that the RMT 
probability distribution (\ref{RMTp}) is not sufficient to describe 
quasi-one-dimensional conductors with transverse structure. The 
generalization of RMT to such conductors remains a challenging problem
\cite{Slevin91a}. 
Beenakker \cite{Beenakker94a} has shown that a correct description of 
quasi-one-dimensional wires without transverse structure requires a 
modification of the interaction in the Hamiltonian (\ref{RMTHam}). It is 
not clear whether a modification of the interaction is 
sufficient to describe conductors with transverse structure or if three 
and more eigenvalue interactions are needed. Since any generalization 
must be consistent with the form (\ref{limdistr1}) it is of considerable 
interest to have explicit results for $\langle\Gamma_{m}\rangle$ and 
${\cal A}_{mn}$ which go beyond RMT and the DMPK equation. Numerical 
simulations \cite{Markos93a,Markos94a} showed that the correlations 
between a pair of $\alpha_{m}$ are rather weak. So it might be that the 
correlations vanish in the thermodynamic limit leading to a diagonal 
form of ${\cal A}_{mn}$. A perturbative calculation of the limiting 
distribution for strong forward scattering will be published in a 
subsequent paper \cite{Endesfelder96c}. It will shed some light on these
questions.
\section*{Acknowledgements}
I would like to thank J. Chalker for many stimulating discussions and a 
critical reading of the manuscript. This work has been supported by the 
Human Capital and Mobility program of the European Union. 
\begin{appendix}
\section{Derivation of the Langevin equations}
\label{ALvgl}
The Langevin equations for $\bbox{\Gamma}$ and $\bbox{u}$ which describe 
the stochastic evolution of the matrix 
\begin{eqnarray}
\bbox{TT}^\dagger\equiv\bbox{M}   
& = & \left(\begin{array}{cc} 
      \bbox{M}^{11} & \bbox{M}^{12} \\ 
      \bbox{M}^{21} & \bbox{M}^{22} \end{array}\right)
\nonumber \\
& = & \left(\begin{array}{cc} 
      \bbox{u}(\cosh\bbox{\Gamma})\bbox{u}^{\dagger} & 
              \bbox{u}(\sinh\bbox{\Gamma})\bbox{u}^{T} \\
      \bbox{u}^{*}(\sinh\bbox{\Gamma})\bbox{u}^{\dagger} & 
              \bbox{u}^{*}(\cosh\bbox{\Gamma})\bbox{u}^{T} 
      \end{array}\right)
\end{eqnarray}
were already given by Dorokhov in his pioneering work 
\cite{Dorokhov83a}. Since he did not derive them explicitly we derive 
them in this appendix. The multiplicative nature 
(\ref{Tmultiplicativity}) of the transfer matrix implies that 
\begin{equation}
\label{ML+dtL}
\bbox{M}(L+\delta L)= \bbox{T}(L+\delta L,L) \bbox{M}(L)
                             \bbox{T}^\dagger (L+\delta L,L)
\end{equation}
if a short segment of length $\delta L$ is added to a sample of length
$L$. The change $\Delta\bbox{M}=\bbox{M}(L+\delta L)-\bbox{M}(L)$ of 
$\bbox{M}$ 
induces the changes $\Delta\bbox{\Gamma}$ and $\Delta\bbox{u}$ which can 
be calculated by perturbation theory. The eigenvalues of the hermitean 
matrix $\bbox{M}^{11}=\bbox{u}\cosh\bbox{\Gamma}\;\bbox{u}^\dagger$ 
are $\cosh\Gamma_m$. The corresponding eigenvector to $\cosh\Gamma_{m}$ 
is the $m^{th}$ column vector $\vec{u}_{m}$of $\bbox{u}$. The change 
$\Delta\bbox{u}$ may be expanded into these eigenvectors
\begin{equation}
\label{Entw}
\Delta \vec{u}_{m}=\sum_{k=1}^N c_{mk} \vec{u}_{k} .
\end{equation}
Non-degenerate first-order perturbation theory then yields
\begin{equation}
\Delta\Gamma_m = \frac{\{\bbox{u}^\dagger\Delta\bbox{M}^{11}\bbox{u}\}_{
mm}}{\sinh\Gamma_m} + \makebox{O} ((\Delta\bbox{M}^{11})^{2})
\end{equation}
and
\begin{equation}
c_{mn}= \frac{\{\bbox{u}^\dagger\Delta\bbox{M}^{11}\bbox{u}\}_{nm}}{
\cosh\Gamma_m-\cosh\Gamma_n} + \makebox{O} ((\Delta\bbox{M}^{11})^{2})
\end{equation}
for $n\ne m$. The expansion coefficient $c_{mm}$ can be calculated from  
\begin{equation}
\label{DtM12}
\Delta\bbox{M}^{12}=\Delta\bbox{u}(\bbox{F}+\Delta\bbox{F})\bbox{u}^T+
\bbox{u}(\bbox{F}+\Delta\bbox{F})\Delta\bbox{u}^T+\bbox{u}\Delta\bbox{F}
\bbox{u}^T+\Delta\bbox{u}(\bbox{F}+\Delta\bbox{F})\Delta\bbox{u}^T
\end{equation}
where $\bbox{F}=\sinh\bbox{\Gamma}$. Equation (\ref{Entw}) implies that 
$\bbox{u}^\dagger\Delta\bbox{u}=\bbox{c}^T$. Multiplying Eq. 
(\ref{DtM12}) with $\bbox{u}^\dagger$ from the left and with $\bbox{u}^*
$ from the right thus yields
\begin{equation}
\bbox{u}^\dagger\Delta\bbox{M}^{12}\bbox{u}^*=
\bbox{c}^T 
(\bbox{F}+\Delta\bbox{F})+(\bbox{F}+\Delta\bbox{F})\bbox{c}+\Delta
\bbox{F}+\bbox{c}^T (\bbox{F}+\Delta\bbox{F})\bbox{c}\hspace{0.5cm} .
\end{equation}                        
Hence 
\begin{equation}
c_{mm}=\frac{\{\bbox{u}^\dagger\Delta\bbox{M}^{12}\bbox{u}^*\}_{mm}-
\coth\Gamma_m
\{\bbox{u}^\dagger\Delta\bbox{M}^{11}\bbox{u}\}_{mm}}{\sinh\Gamma_m}
+\makebox{O} ((\Delta\bbox{M}^{12})^{2})\hspace{0.5cm} .
\end{equation}
Inserting the expansion (\ref{T3expansion}) of 
$\bbox{T}(L+\delta L,L)$ into powers of $\delta L$ into Eq. 
(\ref{ML+dtL}) gives 
\begin{equation}
\label{EntDtM}
\Delta\bbox{M}^{ij} = 
\left(\bbox{\gamma}^{ik}\bbox{M}^{kj}+
      \bbox{M}^{ik}\bbox{\gamma}^{kj\:\dagger}\right) \delta L+
\makebox{O}(\delta L^{2}).                  
\end{equation}
Collecting results and taking the limit $\delta L\to 0$ finally leads to
the Langevin equations
\begin{eqnarray}
\label{Lvgl2}
\frac{d\Gamma_{m}}{dL} & = & E_{mm}+E^{*}_{mm} \nonumber \\[1.5ex]
\frac{d u_{mn}}{dL} & = & \sum_{k} \gamma^{11}_{mk}u_{kn}+
\sum_{k\ne n}\left(\frac{E_{kn}\sinh\Gamma_{n}+E_{kn}^{*}\sinh
\Gamma_{k}}{\cosh\Gamma_{n}-\cosh\Gamma_{k}}\right) u_{mk}\nonumber \\
&  & +\left(\frac{1}{2}\coth\Gamma_{n}(E_{nn}-E^{*}_{nn})\right) u_{mn},
\end{eqnarray}
where $\bbox{E}=\bbox{u}^{+}\bbox{\gamma}^{12}\bbox{u}^{*}$. Note that
the symmetries $\bbox{\gamma}^{11\:\dagger}=-\bbox{\gamma}^{11}$ 
and $\bbox{\gamma}^{12\: T}=\bbox{\gamma}^{12}$ imply that 
$d(\bbox{uu}^{\dagger})/dL|_{\bbox{uu}^{\dagger}=\bbox{1}}=0$
which ensures that unitarity is conserved.
\section{The Coefficients of the FP operator}
\label{AFPkoeff}
The coefficients of the FP operator can be calculated  by iterative 
integration of the Langevin equations (\ref{Lvgl1}) and subsequent 
averaging over the disorder \cite{Risken89}. Integration of the Langevin 
equation yields 
\begin{eqnarray}
\label{Int}
\Gamma_{m}(x) & = & \Gamma_{m}(L)+\int_{L}^{x} dx'(E_{mm}(x')+E^{*}_{mm}
(x')) \nonumber\\[1.5ex]
u_{mn}(x) & = & u_{mn}(L)+\int_{L}^{x} dx' \{\bbox{\gamma}^{11}(x')
\bbox{u}(x')+\bbox{u}(x')\bbox{P}(x')\}_{mn}.  
\end{eqnarray} 
The matrix elements of $\bbox{u}(x')$ and $\bbox{u}^{*}(x')$ which 
appear in the integrands can again be expressed by the second of the Eqs. 
(\ref{Int}). Iterating this procedure leads to an increasing number of 
terms with polynomials of $\bbox{\gamma}^{ij}$ of increasing degree 
$r$. For the Gaussian white noise model (\ref{Mod3}) only polynomials
of even degree have non-zero disorder averages. The integration over the 
$\delta$-functions of the disorder averaged polynomials then leads to 
terms of order $\delta L^{r/2}$. Hence, only polynomials of degree two 
contribute to the limit 
$[\cdots]\equiv \lim_{\delta L\to 0}\langle\cdots\rangle_{\delta L}/
\delta L$. Collecting these polynomials leads to the following result 
for the coefficients of the  FP operator 
\begin{eqnarray}
\label{FPkoeff}
[\Delta\Gamma_{m}] & = & 
[\int_{L}^{L+\delta L} dx \int_{L}^{L+\delta L} dx'\{\bbox{u}^{\dagger}
\bbox{\gamma}^{12}(x)\bbox{\gamma}^{11\: *}(x')\bbox{u}^{*}+ 
\bbox{\cal E}(x)\bbox{\cal P}^{*}(x')+c.c.\}_{mm}]\nonumber \\[2ex]
\left[\Delta\Gamma_{m}\Delta\Gamma_{n}\right]
& = & [\int_{L}^{L+\delta L} dx \int_{L}^{L+\delta L} dx'\{\bbox{\cal E}
(x)+\bbox{\cal E}^{*}(x)\}_{mm}\{\bbox{\cal E}(x')+\bbox{\cal E}^{*}(x')
\}_{nn}]\nonumber \\[2ex]
\left[\Delta\Gamma_{m}\Delta u_{m'n'}\right]
& = & [\int_{L}^{L+\delta L} dx \int_{L}^{L+\delta L} dx' \{\bbox{\cal 
E}(x)+\bbox{\cal E}^{*}(x)\}_{mm}\{\bbox{\gamma}^{11}(x')\bbox{u}+
\bbox{u}\bbox{\cal P}(x'))\}_{m'n'}] \nonumber \\[2ex]
\left[\Delta u_{mn}\right]
& = & \frac{1}{2} [\int_{L}^{L+\delta L} dx \int_{L}^{L+\delta L} dx' \{
\bbox{\gamma}^{11}(x)\bbox{\gamma}^{11}(x')\bbox{u}+2\bbox{\gamma}^{11}
(x)\bbox{u}\bbox{\cal P}(x')\nonumber \\
& &+\bbox{u}\bbox{\cal P}(x')\bbox{\cal P}(x)\}_{mn} \nonumber \\
& & +\sum_j u_{mj}\theta (n-j) \{\bbox{u}^{\dagger}\bbox{\gamma}^{11\:
\dagger}(x')\bbox{\gamma}^{12}(x)\bbox{u}^{*} \nonumber \\
& & +\bbox{u}^{\dagger}\bbox{\gamma}^{12}(x)\bbox{\gamma}^{11\: *}(x')
\bbox{u}^{*}+\bbox{\cal P}^{\dagger}(x')\bbox{\cal E}(x)+\bbox{\cal E}
(x)\bbox{\cal P}^{*}(x')\}_{jn} \nonumber \\
& & -\sum_j u_{mj}\theta (j-n) \{\bbox{u}^{\dagger}\bbox{\gamma}^{11\:
\dagger}(x')\bbox{\gamma}^{12}(x)\bbox{u}^{*} \nonumber \\
& & +\bbox{u}^{\dagger}\bbox{\gamma}^{12}(x)\bbox{\gamma}^{11\: *}(x')
\bbox{u}^{*}+\bbox{\cal P}^{\dagger}(x')\bbox{\cal E}(x)+\bbox{\cal E}
(x)\bbox{\cal P}^{*}(x')\}^{*}_{jn}] \nonumber \\[2ex]
\left[\Delta u_{mn}\Delta u_{m'n'}\right]
& = & [\int_{L}^{L+\delta L} dx \int_{L}^{L+\delta L} dx'
\{\bbox{\gamma}^{11}(x)\bbox{u}+\bbox{u}\bbox{\cal P}(x))\}_{mn}
\nonumber \\
& & \{\bbox{\gamma}^{11}(x')\bbox{u}+\bbox{u}\bbox{\cal P}(x'))
\}_{m'n'}] \nonumber \\[2ex]
\left[\Delta u_{mn}\Delta u^{*}_{m'n'}\right]
& = & [\int_{L}^{L+\delta L} dx \int_{L}^{L+\delta L} dx'
\{\bbox{\gamma}^{11}(x)\bbox{u}+\bbox{u}\bbox{\cal P}(x))\}_{mn}
\nonumber \\
& & \{\bbox{\gamma}^{11\: *}(x')\bbox{u}^{*}+\bbox{u}^{*}\bbox{\cal 
P}^{*}(x'))\}_{m'n'}] \nonumber \\
& &
\end{eqnarray}
where
\begin{eqnarray}
\bbox{\cal E}(x) & = & \bbox{u}^\dagger \bbox{\gamma}^{12}(x)
\bbox{u}^* \nonumber \\[1.5ex]
{\cal P}_{mn}(x) & = & \theta (n-m) {\cal E}_{mn}(x)-\theta (m-n) {\cal 
E}^*_{mn}(x) 
\end{eqnarray}
and integrals of the type $\int_{L}^{L+\delta L} dx \int_{L}^{x} dx' 
\delta (x-x') f(x')$ have been replaced by $\int_{L}^{L+\delta L} dx$ $
\int_{L}^{L+\delta L} dx' \delta (x-x') f(x')/2$.   
\section{The invariant measure of the unitary group}
\label{AInvMeas}
The invariant measure $d\mu(\bbox{u})$ of the unitary group is invariant 
under multiplication with an arbitrary element $\bbox{u}_{0}$ of the 
group from the left and the right
\begin{equation}
d\mu(\bbox{u}) = d\mu(\bbox{u}_{0}\bbox{u}) = d\mu(\bbox{uu}_{0}) .
\end{equation}
As claimed in section \ref{SLvFPgl} we show in this appendix that the 
invariant measure has the form
\begin{eqnarray}
d\mu(\bbox{u}) & = & {\cal V}^{-1}(N) \delta(\bbox{1},\bbox{uu}^{\dagger
})\prod_{m,n}dx_{mn}dy_{mn} ,
\end{eqnarray}
where $u_{mn}=x_{mn}+iy_{mn}$, ${\cal V}(N)$ is the volume of the 
unitary group, and the $\delta$-function $\delta(\bbox{1},\bbox{uu}^{
\dagger})$ has been defined in Eq. (\ref{deltafunction}). Since the 
unitary group is compact, invariance under multiplication from the right 
implies invariance under multiplication from the left (cf. p. $316$  in 
\cite{Hammermesh62}). Therefore it is sufficient to show the right 
invariance. Let us write $\bbox{u}'=\bbox{uu}_0$. Then
\begin{eqnarray}
d\mu(\bbox{u}') & = & {\cal V}^{-1}(N) \delta(\bbox{1},\bbox{u}'\bbox{u
}^{'\:\dagger})\prod_{m,n}dx'_{mn} dy'_{mn} \nonumber \\
& = & {\cal V}^{-1}(N) \delta(\bbox{1},\bbox{u}\bbox{u}^{\dagger})\left|
\det\frac{\partial (x_{m'n'},y_{m'n'})} {\partial (x_{mn},y_{mn})}
\right| \prod_{m,n}dx_{mn}dy_{mn} . 
\end{eqnarray}
Thus $d\mu(\bbox{u})$ is right invariant, if the absolute value of the 
Jacobi determinant of the linear transformation $\bbox{u}'=\bbox{uu}_{
0}$ equals one. This transformation is equivalent to   
\begin{eqnarray}
\left(\begin{array}{c}\vec{\bbox{x}}'\\ \vec{\bbox{y}}'\end{array}
\right) 
& = & 
\left(\begin{array}{cc}
\bbox{1}_{N}\otimes\bbox{x}_{0}^{T} & -\bbox{1}_{N}\otimes\bbox{y}_{0
}^{T} \\ 
\bbox{1}_{N}\otimes\bbox{y}_{0}^{T} & \bbox{1}_{N}\otimes\bbox{x}_{0
}^{T}
\end{array}\right)
\left(\begin{array}{c}\vec{\bbox{x}}\\ \vec{\bbox{y}}\end{array}\right)
\nonumber \\[1.5ex]
& = &  
\frac{\partial (\vec{\bbox{x}}',\vec{\bbox{y}}')}{\partial (\vec{\bbox{x
}},\vec{\bbox{y}})} 
\left(\begin{array}{c}\vec{\bbox{x}}\\ 
                      \vec{\bbox{y}}\end{array}\right),
\end{eqnarray}
where $\bbox{u}=\bbox{x}+i\bbox{y}$ has been written in the vector form
\begin{equation}
\left(\begin{array}{cc}
\vec{\bbox{x}}^{T} & \vec{\bbox{y}}^{T}
\end{array}\right)
=\left(\begin{array}{cccccccccc} 
x_{11} & x_{12} & \cdots & x_{1N} & x_{21} & \cdots & x_{NN} & y_{11} & 
\cdots & y_{NN} \end{array}\right) .
\end{equation}
The multiplication rule $(\bbox{A}\otimes\bbox{B})(\bbox{C}\otimes\bbox{
D})=(\bbox{A}\bbox{C})\otimes(\bbox{B}\bbox{D})$ for cross products of 
matrices and the unitarity of $\bbox{u}$ imply that $\partial (\vec{
\bbox{x}}',\vec{\bbox{y}}')/ \partial (\vec{\bbox{x}},\vec{\bbox{y}})$ 
is an orthogonal matrix. Therefore the absolute value of the Jacobi 
determinant is one and the measure is right invariant.
\section{Alternative derivation of the FP equation}
\label{AaltderFP}
The FP equation (\ref{FPgl1}) can be derived in an alternative way 
yielding immediately the form in which the derivatives act directly on 
the probability distribution and not on the coefficients of the FP 
operator. This implies useful relations between these two forms which 
are difficult to obtain by direct differentiation of the coefficients if
the result is not known in advance. Therefore we describe the 
alternative derivation in this appendix. It is a generalization of the 
technique which was employed by Mello and coworkers to derive the FP 
equation of their isotropic model \cite{Mello87a,Mello88c}. 

Assume that the probability distribution of  $\bbox{M}$ for a system of 
length $L$ is $\tilde{p}(L;\bbox{M})$. Then add a statistically 
independent segment of length $\delta L$ to the system. The distribution 
of the 
transfer matrix $\bbox{T}_{\delta L}$ of the segment is denoted by $w(L,
\delta L;\bbox{T}_{\delta L})$. Averaging a function $f(\bbox{M})$ 
over the disorder of the whole system yields 
\begin{eqnarray}
\label{Convolution1}
\langle f(\bbox{M})\rangle_{L+\delta L}  
& \equiv & \int d\rho\: (\bbox{M})\;\tilde{p}(L+\delta L;\bbox{M})
\;f(\bbox{M})\nonumber\\ 
& = & \int\int d\rho\: (\bbox{M}')\; d\omega\: (\bbox{T}_{\delta L})\; 
\tilde{p}(L;\bbox{M}')\; w(L,\delta L;\bbox{T}_{\delta L}) 
f(\bbox{T}_{\delta L}\bbox{M}'\: \bbox{T}_{\delta L}^\dagger)
\end{eqnarray}
where  $\bbox{M}=\bbox{T}_{\delta L}\bbox{M}'\:\bbox{T}_{\delta L}^
\dagger$ and $d\omega\: (\bbox{T})$, $d\rho\: (\bbox{M})$ are measures 
on the matrix spaces of $\bbox{T}$ and $\bbox{M}$. If the measure $d\rho
\: (\bbox{M})$ is chosen to be invariant under the transformation 
$\bbox{T}_{0}\bbox{M}\:\bbox{T}^{\dagger}_{0}$ for any transfer matrix 
$\bbox{T}_{0}$ one finds 
\begin{eqnarray}
\label{Convolution2}
\langle f(\bbox{M})\rangle_{L+\delta L}
& = &\int\int d\rho\: (\bbox{M})\; d\omega\: (\bbox{T}_{\delta L})\;
     \tilde{p}(L;\bbox{T}_{\delta L}^{-1}\bbox{M}(\bbox{T}_{\delta 
     L}^\dagger)^{-1}) w(L,\delta L;\bbox{T}_{\delta L})\; f(\bbox{M}).
\end{eqnarray}
A similar line of reasoning as in \cite{Mello88c} yields that the 
invariant measure has the form 
\begin{equation}
d\rho\: (\bbox{M}) = J(\bbox{\Gamma})\prod_{m=1}^{N} d\Gamma_m d\mu (
\bbox{u})
\end{equation} 
where $J(\bbox{\Gamma})=\prod_{m<n} |\cosh\Gamma_m -\cosh\Gamma_n|
\prod_m\sinh\Gamma_m$ and $d\mu (\bbox{u})$ is the invariant measure on
the unitary group. Comparing Eq. (\ref{Convolution1}) with Eq. 
(\ref{Convolution2}) immediately shows that
\begin{equation}     
\label{Convolution3}
\tilde{p}(L+\delta L;\bbox{M})=\int d\omega\: (\bbox{T}_{\delta L})\;
\tilde{p}(L;\bbox{T}_{\delta L}^{-1} \bbox{M}\,(\bbox{T}_{\delta 
L}^\dagger)^{-1})\; w(L,\delta L;\bbox{T}_{\delta L})
\end{equation}
or equivalently
\begin{equation} 
\label{Convolution4}
\tilde{p}(L+\delta L;\bbox{M})=\langle \tilde{p}(L;\bbox{T}_{\delta L}^{
-1}\bbox{M}\,(\bbox{T}_{\delta L}^\dagger)^{-1})\rangle_{\delta L}
\end{equation}
since the average of $\tilde{p}(L;\bbox{T}_{\delta L}^{-1}
\bbox{M}\,(\bbox{T}_{\delta L}^\dagger)^{-1})$ with the probability 
$w(L,\delta L;\bbox{T}_{\delta L})d\omega (\bbox{T}_{\delta L})$ is just 
the average $\langle\cdots\rangle_{\delta L}$ over the disorder of the 
segment.
For a fixed value of $\bbox{\Gamma}$ the distribution $\tilde{p}(L+
\delta L;\bbox{M})$ may be expanded into the irreducible representations
of the unitary group. Since the irreducible representations are 
polynomials  in $u_{mn}$ and $u_{mn}^{*}$ \cite{Weyl46,Louck70a}, 
$\tilde{p}(L+\delta L;\bbox{M})$ can be analytically continued to 
arbitrary complex matrix elements $u_{mn}$. This justifies to write Eq. 
(\ref{Convolution4}) in the form 
\begin{equation}
\label{Convolution5}
\tilde{p}(L+\delta L;\bbox{\Gamma},\bbox{u},\bbox{u}^*)=
\langle \tilde{p}(L;\bbox{\Gamma}+\Delta \bbox{\Gamma},\bbox{u}+\Delta
\bbox{u},\bbox{u}^*+\Delta \bbox{u}^*)\rangle_{\delta L} ,
\end{equation}
where $\Delta\bbox{\Gamma}$, $\Delta\bbox{u}$ and $\Delta\bbox{u}^{*}$ 
are the changes of the parameters of $\bbox{M}$ which are induced by the
transformation $\bbox{T}_{\delta L}^{-1} \bbox{M}\,(\bbox{T}_{\delta L}^
\dagger)^{-1}$. A Taylor expansion of the left side in powers of $\delta
L$ and of the right side in powers of $\Delta\Gamma_{m}$, $\Delta u_{mn}
$ and $\Delta u^{*}_{mn}$ yields
\begin{eqnarray}
\label{FPgl4}
\frac{\partial\tilde{p}}{\partial L} & = & 
\left\{\frac{1}{2} [\Delta\Gamma_m\Delta\Gamma_{n}]\partial_{\Gamma_{m}}
\partial_{\Gamma_{n}}+[\Delta\Gamma_m]\partial_{\Gamma_{m}}+[\Delta
\Gamma_m\Delta u_{m'n'}]\partial_{\Gamma_{m}}\partial_{u_{m'n'}}\right.
\nonumber \\
&  &\hspace*{2.7mm} +[\Delta\Gamma_m\Delta u^{*}_{m'n'}]\partial_{
\Gamma_{m}}\partial_{u^{*}_{m'n'}}+[ \Delta u_{mn}] \partial_{u_{mn}}+ 
[\Delta u_{mn}^*]\partial_{u_{mn}^*} \nonumber \\
&  &\hspace*{2.7mm} +[\Delta u_{mn}\Delta u_{m'n'}^*] \partial_{u_{mn}}
\partial_{u_{m'n'}^*}+ \frac{1}{2}[\Delta u_{mn}\Delta u_{m'n'}]
\partial_{u_{mn}}\partial_{u_{m'n'}}\nonumber \\ 
&  &\hspace*{2.7mm} \left. +\frac{1}{2}[\Delta u_{mn}^* \Delta u_{m'n'
}^*]\partial_{u_{mn}^*} \partial_{u_{m'n'}^*}\right\}\tilde{p}
\end{eqnarray}       
in the limit $\delta L\to 0$. The coefficients $[\cdots]\equiv\lim_{
\delta L\to 0}\langle\cdots\rangle_{\delta L}/\delta L$ could be 
determined by the calculation of  $\Delta\Gamma_{m}$, $\Delta u_{mn}$ 
and $\Delta u^{*}_{mn}$ with perturbation theory and subsequent 
averaging over the disorder. This would lead to expressions which 
involve $[\bar{\varepsilon}^{ij}_{mn}]$ and $[\bar{\varepsilon}^{
ij}_{mn}\bar{\varepsilon}^{i'j'}_{m'n'}]$ where 
\begin{equation}
\bbox{T}(L+\delta L,L)^{-1} = 
\bbox{1}+\left(
\begin{array}{ll} 
\bar{\bbox{\varepsilon}}^{11} & \bar{\bbox{\varepsilon}}^{12} \\
\bar{\bbox{\varepsilon}}^{21} & \bar{\bbox{\varepsilon}}^{22}
\end{array}\right),
\end{equation} 
$\bar{\bbox{\varepsilon}}^{11}$=$\bbox{\varepsilon}^{11\:\dagger}$, and
$\bar{\bbox{\varepsilon}}^{12}=-\bbox{\varepsilon}^{12\: T}$. In order
to include all the terms which contribute to the coefficients
one had to go to the second order of the perturbation
theory which is quite involved. We proceed in a different way, instead.
For the Gaussian white noise model (\ref{Mod3}) one finds  
\begin{eqnarray}
[\bar{\varepsilon}^{ij}_{mn}] & = & [\varepsilon^{ij}_{mn}] 
\nonumber\\[1ex]
[\bar{\varepsilon}^{ij}_{mn}\bar{\varepsilon}^{i'j'}_{m'n'}] & = & 
[\varepsilon^{ij}_{mn}\varepsilon^{i'j'}_{m'n'}].
\end{eqnarray}
Hence, the coefficients may be as well evaluated with the changes 
$\Delta\Gamma_{m}$, $\Delta u_{mn}$ and $\Delta u^{*}_{mn}$ which 
are induced by the transformation $\bbox{T}_{\delta L}
\bbox{M}\bbox{T}_{\delta L}^\dagger$. These changes can be obtained by 
iterative integration of the Langevin equations (\ref{Lvgl2}) similar
to that described in appendix \ref{AFPkoeff} for the simpler case of 
large system lengths.
 
Since $\mbox{d}\rho(\bbox{M})/J(\bbox{\Gamma})$=$\prod_{m}d\Gamma_{m}
d\mu(\bbox{u})$ is the same measure which was used for the probability
distribution of the FP equation (\ref{FPgl1}), one expects
that the distribution $J(\bbox{\Gamma})\tilde{p}(L;\bbox{\Gamma},
\bbox{u},\bbox{u}^*)$ obeys this FP equation for large system lengths. 
In fact, the transformation  of Eq. (\ref{FPgl4}) to this 
distribution and the neglect of the exponential small contributions to
terms of the form (\ref{approx}) leads to Eq. (\ref{FPgl1}) 
where the operators $\hat{A}$, $\hat{B}$ and $\hat{C}$ appear in the 
form
\begin{eqnarray}
\label{FPOp2}
\hat{A}_{mn} & = & \frac{1}{2} \left[ \Delta\Gamma_m \Delta\Gamma_n 
\right]\nonumber \\[1.5ex]
\hat{B}_m & = & \left[ \Delta\Gamma_m \right] - n\left[ \Delta\Gamma_n 
\Delta\Gamma_m \right] \nonumber \\ 
&   & + \left[\Delta\Gamma_m\Delta u_{m'n'}\right] \partial_{u_{m'n'}} +
 \left[\Delta\Gamma_m\Delta u_{m'n'}^*\right] \partial_{u_{m'n'}^*} 
\nonumber \\[1.5ex]
\hat{C} & = & \frac{1}{2} m n\left[ \Delta\Gamma_m \Delta\Gamma_n 
\right]-m\left[ \Delta\Gamma_m \right] -  m\left[\Delta\Gamma_m\Delta 
u_{m'n'}\right]\partial_{u_{m'n'}} \nonumber \\
&  &  - m\left[\Delta\Gamma_m\Delta u_{m'n'}^*\right]  \partial_{
u_{m'n'}^*}+\left[ \Delta u_{mn}\right]\partial_{u_{mn}}+ \left[ \Delta 
u_{mn}^*\right]\partial_{u_{mn}^*} \nonumber \\ 
& & +\left[ \Delta u_{mn} \Delta u_{m'n'}^* \right]\partial_{u_{mn}} 
\partial_{u_{m'n'}^*} + \frac{1}{2} \left[ \Delta u_{mn} \Delta u_{m'n'} 
\right] \partial_{u_{mn}}  \partial_{u_{m'n'}}  \nonumber \\ 
& & +\frac{1}{2} \left[ \Delta u_{mn}^* \Delta u_{m'n'}^* \right] 
\partial_{u_{mn}^*}\partial_{u_{m'n'}^*} .
\end{eqnarray}
The equivalence with the form of $\hat{A}$, $\hat{B}$ and $\hat{C}$ in 
Eq. (\ref{FPOp1}) can be shown by a lengthy but straightforward 
calculation  which exploits only the symmetries $\bbox{\gamma}^{11\:
\dagger}=-\bbox{\gamma}^{11}$ and $\bbox{\gamma}^{12\: T}=\bbox{\gamma
}^{12}$.  
\end{appendix}


\begin{thebibliography}{10}

\bibitem{MesoscopicRev}
 For reviews, see {\em Mesoscopic Phenomena in Solids}, edited by 
 B.~L. Altshuler, P.~A. Lee, and R.~A. Webb 
 (North-Holland, Amsterdam, 1991) and
 {\em Quantum Coherence in Mesoscopic Systems}, edited by B. Kramer (NATO
  ASI Series B, vol. 254, New York: Plenum, 1991).

\bibitem{Buettiker90a}
M. B\"u{}ttiker, Phys.\ Rev.\ Lett. {\bf 65},  2901  (1990).

\bibitem{Beenakker91a}
C.~W.~J. Beenakker, Phys.\ Rev.\ Lett. {\bf 67},  3836  (1991).

\bibitem{Beenakker92a}
C.~W.~J. Beenakker, Phys.\ Rev.\ B {\bf 46},  12841  (1992).

\bibitem{Imry86a}
Y. Imry, Europhys.\ Lett. {\bf 1},  249  (1986).

\bibitem{Muttalib87a}
K.~A. Muttalib, J.-L. Pichard, and A.~D. Stone, Phys.\ Rev.\ Lett. {\bf 59},
  2475  (1987).

\bibitem{Beenakker93c}
C.~W.~J. Beenakker, Phys.\ Rev.\ Lett. {\bf 70},  1155  (1993).

\bibitem{Beenakker93a}
C.~W.~J. Beenakker, Phys.\ Rev.\ B {\bf 47},  15763  (1993).

\bibitem{Mello88a}
P.~A. Mello, Phys.\ Rev.\ Lett. {\bf 60},  1089  (1988).

\bibitem{Mello88c}
P.~A. Mello, P. Pereyra, and N. Kumar, Ann. Phys. (N.Y.) {\bf 181},  290
  (1988).

\bibitem{Hueffmann90a}
A. H\"u{}ffmann, J. Phys. A {\bf 23},  5733  (1990).

\bibitem{Mello91b}
P.~A. Mello and A.~D. Stone, Phys.\ Rev.\ B {\bf 44},  3559  (1991).

\bibitem{Macedo92a}
A.~M.~S. Mac\^e{}do and J.~T. Chalker, Phys.\ Rev.\ B {\bf 46},  3559  (1991).

\bibitem{Iida90b}
S. Iida, H.~A. Weidenm\"u{}ller, and J.~A. Zuk, Phys.\ Rev.\ Lett. {\bf 64},
  583  (1990).

\bibitem{Iida90a}
S. Iida, H.~A. Weidenm\"u{}ller, and J.~A. Zuk, Ann. Phys. {\bf 200},  219
  (1990).

\bibitem{Altland91a}
A. Altland, Z. Phys. B {\bf 82},  105  (1991).

\bibitem{Lee85a}
P.~A. Lee and A.~D. Stone, Phys.\ Rev.\ Lett. {\bf 55},  1622  (1985).

\bibitem{Altshuler85a}
B.~L. Al'tshuler, Pis'ma Zh. Eksp. Teor. Fiz. {\bf 41},  530  (1985) 
[JETP Lett. {\bf 41},  648  (1985)].

\bibitem{Chalker93c}
J.~T. Chalker and A.~M.~S. Mac\^e{}do, Phys.\ Rev.\ Lett. {\bf 71},  3693
  (1993).

\bibitem{Macedo94a}
A.~M.~S. Mac\^e{}do and J.~T. Chalker, Phys.\ Rev.\ B {\bf 49},  4695  (1994).

\bibitem{Beenakker93b}
C.~W.~J. Beenakker and B. Rejaei, Phys.\ Rev.\ Lett. {\bf 71},  3689  (1993).

\bibitem{Beenakker94a}
C.~W.~J. Beenakker and B. Rejaei, Phys.\ Rev.\ B {\bf 49},  7499  (1994).

\bibitem{Caselle95a}
M. Caselle, Phys.\ Rev.\ Lett. {\bf 74},  2776  (1995).

\bibitem{Zirnbauer92a}
M.~R. Zirnbauer, Phys.\ Rev.\ Lett. {\bf 69},  1584  (1992).

\bibitem{Mirlin94a}
A.~D. Mirlin, A. M\"u{}ller-Groeling, and M.~R. Zirnbauer, Ann. Phys. {\bf
  236},  325  (1994).

\bibitem{Frahm95a}
K. Frahm, Phys.\ Rev.\ Lett. {\bf 74},  4706  (1995).

\bibitem{Mello91c}
P.~A. Mello and S. Tomsovic, Phys.\ Rev.\ Lett. {\bf 67},  342  (1991).

\bibitem{Mello92a}
P.~A. Mello and S. Tomsovic, Phys.\ Rev.\ B {\bf 46},  15963  (1992).

\bibitem{Chalker93a}
J.~T. Chalker and M. Bernhardt, Phys.\ Rev.\ Lett. {\bf 70},  982  (1993).

\bibitem{Nazarov94a}
Y.~V. Nazarov, Phys.\ Rev.\ Lett. {\bf 73},  134  (1994).

\bibitem{Dorokhov82a}
O.~N. Dorokhov, Pis'ma Zh. Eksp. Teor. Fiz. {\bf 36},  259  (1982) [JETP 
Lett. {\bf 36}, 318 (1982)].

\bibitem{Dorokhov83a}
O.~N. Dorokhov, Zh. Eksp. Teor. Fiz. {\bf 85},  1040  (1983)  [Sov. Phys. 
JETP {\bf 58},  606  (1983)].

\bibitem{Kree81a}
R. Kree and A. Schmid, Z. Phys. B {\bf 42},  297  (1981).

\bibitem{Endesfelder93a}
D. Endesfelder and B. Kramer, Phys. Rev. E {\bf 48},  R3225  (1993).

\bibitem{Endesfelder96c}
D. Endesfelder, unpublished  .

\bibitem{Mello91a}
P.~A. Mello and J.-L. Pichard, J. Phys. I (France) {\bf 1},  493  (1991).

\bibitem{Oseledec68a}
V.~I. Oseledec, Trans. Moscow Math. Soc. {\bf 19},  197  (1968).

\bibitem{Tutubalin68a}
V.~N. Tutubalin, Theor. Prob. Appl. {\bf 13},  65  (1968).

\bibitem{Virtser70a}
A.~D. Virtser, Theor. Prob. Appl. {\bf 15},  667  (1970).

\bibitem{Johnston83a}
R. Johnston and H. Kunz, J. Phys. C {\bf 16},  3895  (1983).

\bibitem{Cook90a}
J. Cook and B. Derrida, J. Stat. Phys. {\bf 61},  961  (1990).

\bibitem{Derrida87a}
B. Derrida, K. Mecheri, and J.~L. Pichard, J. Phys. (Paris) {\bf 48},  733
  (1987).

\bibitem{Zanon88a}
N. Zanon and B. Derrida, J. Stat. Phys. {\bf 50},  509  (1988).

\bibitem{Markos89a}
P. Marko\v{s}{}, J. Phys.:\ Condens.\ Matter {\bf 1},  4611  (1989).

\bibitem{Markos93c}
P. Marko\v{s}{}, J. Stat. Phys. {\bf 70},  899  (1993).

\bibitem{Markos93a}
P. Marko\v{s}{} and B. Kramer, Ann. Physik {\bf 2},  339  (1993).

\bibitem{Markos94a}
P. Marko\v{s}{}, J. Phys. I France {\bf 4},  551  (1994).

\bibitem{Pichard91a}
J.-L. Pichard,  in {\em Quantum Coherence in Mesoscopic Systems}, edited by B.
  Kramer (NATO ASI Series B, vol. 254, New York: Plenum, 1991).

\bibitem{Pichard90a}
J.-L. Pichard, N. Zanon, Y. Imry, and A.~D. Stone, J. Phys. France {\bf 51},
  587  (1990).

\bibitem{Pichard86a}
J.-L. Pichard and G. Andr\'e{}, Europhys.\ Lett. {\bf 2},  477  (1986).

\bibitem{Weller82a}
W. Weller, V.~N. Prigodin, and Y.~A. Firsov, Phys. Status Solidi B {\bf 110},
  143  (1982).

\bibitem{Apel83a}
W. Apel and T.~M. Rice, J. Phys. C {\bf 16},  L1151  (1983).

\bibitem{Prigodin86a}
V.~N. Prigodin, Y.~A. Firsov, and W. Weller, Solid State Commun. {\bf 59},  729
   (1986).

\bibitem{Weller88a}
W. Weller and M. Kasner, Phys. Status Solidi B {\bf 148},  273  (1988).

\bibitem{Kasner88a}
M. Kasner and W. Weller, Phys. Status Solidi B {\bf 148},  635  (1988).

\bibitem{Wegner79a}
F.~J. Wegner, Phys.\ Rev.\ B {\bf 19},  783  (1979).

\bibitem{Risken89}
H. Risken, {\em The Fokker-Planck Equation} (Springer-Verlag, Berlin
  Heidelberg, 1989).

\bibitem{Slevin91a}
K. Slevin, J.-L. Pichard, and P.~A. Mello, Europhys.\ Lett. {\bf 16},  649
  (1991).

\bibitem{Hammermesh62}
M. Hammermesh, {\em Group Theory and its Applications to Physcics} (Addison,
  Wesley, Reading, Ma, 1962).

\bibitem{Mello87a}
P.~A. Mello, Phys.\ Rev.\ B {\bf 35},  1082  (1987).

\bibitem{Weyl46}
H. Weyl, {\em The Classical Groups} (Princeton University Press, Princeton,
  N.J., 1946).

\bibitem{Louck70a}
J.~D. Louck, Am. J. Phys. {\bf 38},  3  (1970).

\end{thebibliography}
\end{document}